\documentclass[
12pt, 
english, 
doublespacing, 
nolistspacing, 
liststotoc, 
toctotoc, 
parskip, 
headsepline, 
consistentlayout, 
parskip
]{article}
	
\usepackage[utf8]{inputenc}
\usepackage{setspace}
\usepackage[T1]{fontenc}
\usepackage{textcomp}
\usepackage{array}
\usepackage{amsmath}
\usepackage{multirow}
\usepackage{graphicx}
\graphicspath{ {./images/} }
\usepackage{caption}
\usepackage{subcaption}
\usepackage{booktabs}  
\usepackage{siunitx}
\usepackage{footnote}
\usepackage{booktabs,caption}
\usepackage[flushleft]{threeparttable}
\usepackage{longtable}
\usepackage{threeparttablex}
\usepackage{threeparttable} 
\usepackage{booktabs}
\usepackage{threeparttable}
\usepackage{adjustbox}
\usepackage{array}
\usepackage{qrcode}
\usepackage{hyperref}

\usepackage[a4paper, total={6in, 8in}]{geometry}
\usepackage{url}
\usepackage{appendix}
\usepackage{amsbsy}
\usepackage{lscape}
\usepackage{tabularx}
\usepackage{float}
\usepackage[english]{babel}

\usepackage[backend = biber, style = apa, citestyle =authoryear]{biblatex} 
\usepackage[english]{babel}
\usepackage[autostyle=true]{csquotes} 

\usepackage{hyphenat}
\usepackage[dvipsnames]{xcolor}   
\usepackage{hyperref}
\hypersetup{
    colorlinks = true, 
    linktoc = all, 
    linkcolor =red,  
    urlcolor =BurntOrange, 
    citecolor = blue 
}
\usepackage{booktabs}
\usepackage{xcolor,colortbl}
\usepackage{graphics}

\def\frontmatter{%
    \pagenumbering{roman}
    \setcounter{page}{1}
    \renewcommand{\thesection}{\Roman{section}}
}

\def\mainmatter{%
    \pagenumbering{arabic}
    \setcounter{page}{1}
    \setcounter{section}{0}
    \renewcommand{\thesection}{\arabic{section}}
}

\begin{document}

\frontmatter


\begin{titlepage}
\begin{center}




\emph{Author:}\\
\begin{center}
\begin{tabular}{ c c }
P. Carmel Marie Frédérique & ZAGRÉ  \\

\end{tabular}
\end{center}

\vfill
\vfill
\vfill
\vfill

\vfill
\vfill
\vfill
\vfill
\vfill

\vfill
\vfill
\vfill
\vfill
\vfill
\textcopyright \textbf{P. Carmel Marie Frédérique ZAGRÉ 2025} 
\vfill
    \begin{minipage}{\dimexpr\textwidth-1in\relax}
        \small\centering
        The author holds copyright in this work. Please ensure that any reproduction or re-use is done in accordance with the relevant national copyright legislation.
    \end{minipage}
\vfill
\end{center}
\end{titlepage}

\vspace{2cm} 
\newpage
\renewcommand\contentsname{Contents}
\renewcommand\listfigurename{LISTE DES FIGURES}
\renewcommand\listtablename{LISTE DES TABLEAUX}
\renewcommand\bibname{BIBLIOGRAPHIE}
\renewcommand\indexname{INDEX}
\renewcommand\figurename{Figure}
\renewcommand\tablename{Tableau}
\renewcommand\partname{PARTIE}
\renewcommand\appendixname{APPENDICE}
\renewcommand\abstractname{R\'esum\'e}

\newpage
\newpage
\newpage

\newpage
\newpage

\newpage
\mainmatter
\doublespacing
\section{Introduction}

The sustainability of democratic political institutions depends on citizens’ active participation. By voting, respecting others’ rights, obeying laws, and engaging in civic life, citizens uphold the norms and values that sustain democracy. At the same time, the most fundamental duty of any government is to provide security. The state’s capacity to maintain basic security and public order is also contingent on the functioning of democracy. When governments fail to protect citizens from non-state actors, such as jihadist groups, who commit atrocities, individuals may come to prioritize safety and stability over the democratic principles that ensure a flourishing political system.

In recent years, terrorism has resurfaced worldwide, and its presence and impact are more pronounced in Sub-Saharan Africa than ever before. Since 2015, Burkina Faso has been grappling with a severe terrorism crisis that has profoundly disrupted its security and governance structures. In 2024, the country is ranked the most affected by terrorism globally for the first time in over a decade \parencite{InstituteForEconomicsAndPeace2024}. One in every four terrorism-related deaths worldwide now occurs in Burkina Faso \parencite{InstituteForEconomicsAndPeace2024}.  In 2022, the Group for the Support of Islam and Muslims (JNIM) carried out over 400 attacks across 10 of Burkina Faso’s 13 regions, according to the U.S. Department of State (\parencite{USDepartmentofState2023}). 

The escalation of terrorist violence has exposed the fragility of state institutions and undermined progress toward democratic governance. Since 2020, the Sub-Saharan African region has experienced seven coup attempts, four of which succeeded, leading to the rise of authoritarian regimes in Guinea, Mali, Burkina Faso, and, most recently, Niger.  In light of the recent increase in military coups, a notable correlation has been observed: as terrorism grows, democratic governance declines. While one might view democracy as a protective barrier against terrorism, this correlation might also indicate that increased terrorism diminishes support for democratic values. As a result, individuals may begin to favor authoritarian regimes instead. The latter alternative is the focal point of our paper: Does terrorism affect citizens’ support for democracy? Moreover, to what extent does the shift in citizens’ support from democracies to military coups relate to the perceived security threat posed by terrorism? We hypothesize that terrorist groups can indirectly foster democratic backsliding by generating fear among citizens, exposing state incapacity to manage insecurity, and undermining economic conditions and livelihoods.

We combine geocoded terrorism event data with six waves of political attitude surveys to assess how terrorism shapes political attitudes empirically. Specifically, we estimate the dynamic causal effect of successful terrorist attacks on citizens’ support for democracy and military rule over an event window spanning three years before and two years after the a terror attack.  For our identification, we leverage the quasi-random variation in the success of terrorist attacks\footnote{As in \textcite{BrodeurAbel2018TEoT}.} and employ a Staggered Difference-in-Differences (DiD) strategy. Essentially, our approach compares individuals within departments that experienced a successful terrorist attack to those in departments where such attacks were unsuccessful. First, balance tests across a broad range of departmental characteristics show no significant pre-attack differences in social, economic, demographic, geographic, or political covariates between areas that experienced successful and failed attacks. This provides additional support for the validity of our identification strategy.

Second, the analysis reveals a statistically significant relationship between exposure to successful terrorist attacks and citizens' political attitudes. A successful attack increases the probability of supporting military rule by one percentage point and decreases support for democracy by 0.91 percentage points. The event-study plots confirm the dynamic nature of these effects. Before the attack, all pre-treatment coefficients hover around zero, validating the parallel-trends assumption. The impact appear one year after the attack and persists for at least two years, suggesting that terrorist violence does not immediately alter political beliefs but produces lagged and lasting shifts in attitudes. The sensitivity analysis using \textcite{RambachanAshesh2023AMCA} framework further strengthens this interpretation. The treatment effects remain robust to moderate violations of parallel-trend assumptions under both the relative magnitude and smoothness restrictions.

In addition, the results are robust to multiple identification and measurement checks. Controlling for attacks in neighboring departments does not alter the estimated effects, suggesting that the results are not driven by geographic spillovers. Changing the definition of our control group, applying entropy balancing to adjust for compositional differences, and replicating the analysis with a different dataset all produce consistent results. A placebo test based on 700 random permutations of treatment timing further reinforces the causal interpretation, as the distribution of placebo estimates centers tightly around zero.

Third, extending the window allow to examine whether the estimated effects persist over a longer horizon. The direction of the effects remains consistent with our primary findings, but this specification allows us to assess the potential long-term scope of the impact. Overall, whether for support for democracy or for military rule, we observe that the effects tend to fade approximately three years after a successful attack.

Finally, the mechanism analyses reveal that terrorism influences political preferences through three key channels. First, it shifts citizens' priorities from liberty toward security, reflecting a clear security–freedom trade-off. Second, it erodes trust in democratic institutions, particularly the presidency and parliament, while support for the military appears short-lived. Third, terrorism exacerbates local economic hardship, increasing unemployment and perceived poverty. Together, these mechanisms explain how persistent insecurity reshapes citizens' commitment to democracy and fuels temporary support for authoritarian alternatives.

Previous studies examine terrorism's impact on macroeconomic variables such as GDP (\textcite{abadie2003economic}, \textcite{blomberg_macroeconomic_2004}, and \textcite{TAVARES20041039}), trade (\textcite{RePEc:tpr:restat:v:88:y:2006:i:4:p:599-612}; \textcite{deSousaJosé2009TatS}; \textcite{MirzaDaniel2014ALaS}; \textcite{BandyopadhyaySubhayu2018TatA} ), financial markets (\textcite{ChenAndrewH.2004Teot}), tourism (\textcite{EndersWalter1992AEAo}; \textcite{ArinK.Peren2008Tpot}), and Foreign Direct Investment (FDI) (\textcite{AbadieAlberto2008Tatw}). Terrorism has a negative impact on GDP, although the effect is relatively small and short-lived compared to internal or external conflicts. Additionally, it increases transaction costs, uncertainty, and insurance premiums, resulting in declines in trade. In financial markets, terrorism tends to increase risk premiums, making borrowing more expensive. We also observe greater volatility in developing financial markets following terrorist events. Tourism, not surprisingly, experiences declines in both arrivals and revenues following terrorist attacks.

While macroeconomic research highlights terrorism's broad effects on GDP, trade, and investment, more recent studies have focused on its microeconomic consequences, particularly on school enrollment (\textcite{alfano_terrorism_2024} ), labour market ( \textcite{BrodeurAbel2018TEoT}), and firm (\textcite{guo_does_2022}; \textcite{belmonte_selection_2019}, \textcite{FICH2023104655}). The results suggest that terrorism reduces school enrollment, employment among low-skilled workers, and firm capital investments and inventor productivity.

Our research mainly contributes to the strand of literature exploring the impact of terrorism on political attitudes. A growing body of work finds that terrorism can shift political preferences, often toward conservatism and strong leadership. Seminal studies have demonstrated this effect in well-established democracies. For instance, \textcite{AKAY2020103394},  using day-to-day variation in terrorism intensity over 20 years, find that global terrorism reduces subjective well-being and increases the conservative vote share for a subset of western countries. Similarly,  \textcite{PERI2023103864},  using the within-country temporal variation in exposure to fatal terrorist attacks, show a similar shift in Europe. These studies focus on stable Western democracies, neglecting the Sub-Saharan countries and creating a gap in the literature. Many states in these regions of the world face chronic terrorism. 

More recently, \textcite{YameogoNeundorf2025} have begun to address this context gap by analyzing fragile states. Although \textcite{YameogoNeundorf2025} provides evidence by exploiting within-country temporal variation in a single cross-section of the Afrobarometer, our identification strategy extends this approach in two dimensions. First, we leverage the panel structure of multiple survey rounds, which allows us to track the temporal evolution of democratic attitudes within the same regions before and after attacks. This dynamic framework helps us understand how the impact of terrorism spreads and potentially persists over time. Understanding this dynamic is crucial in the context of Burkina Faso, where terrorism has been a persistent threat for nearly a decade. 
Second, our identification hinges on a novel source of quasi-random variation: the success or failure of terrorist attacks, a strategy inspired by \textcite{BrodeurAbel2018TEoT}. By comparing departments that experienced a successful terrorist attack to those that experienced only failed or no attack attempts over time, we isolate the causal effect of an actual attack from that of mere exposure to attempted violence. Finally, our empirical model explicitly controls for attack characteristics and target types to account for potential heterogeneity in terrorists' tactics. By focusing on a critical and understudied region, we provide novel evidence on \textit{how} chronic terrorism reshapes the political landscape in fragile democracies over time. 

The rest of the paper is organized as follows: Section  2 describes the institutional setting of our study, including details on terrorism in Burkina-Faso. In Section 3, we
discuss two main theoretical frameworks that link terrorist attacks to political preference. In Section 4, we provide sources and other relevant details regarding our
data. Section 5 discusses and evaluates our identification strategy. In Section 6, we present our baseline estimating equation and results, while in Section 7, we present evidence on mechanisms that drive our effects. Finally, we conclude in
Section 8.

\section{Political Evolution of Burkina Faso and the rise of terrorism}
\begin{figure}[ht]
    \centering
    \includegraphics[width=1\textwidth]{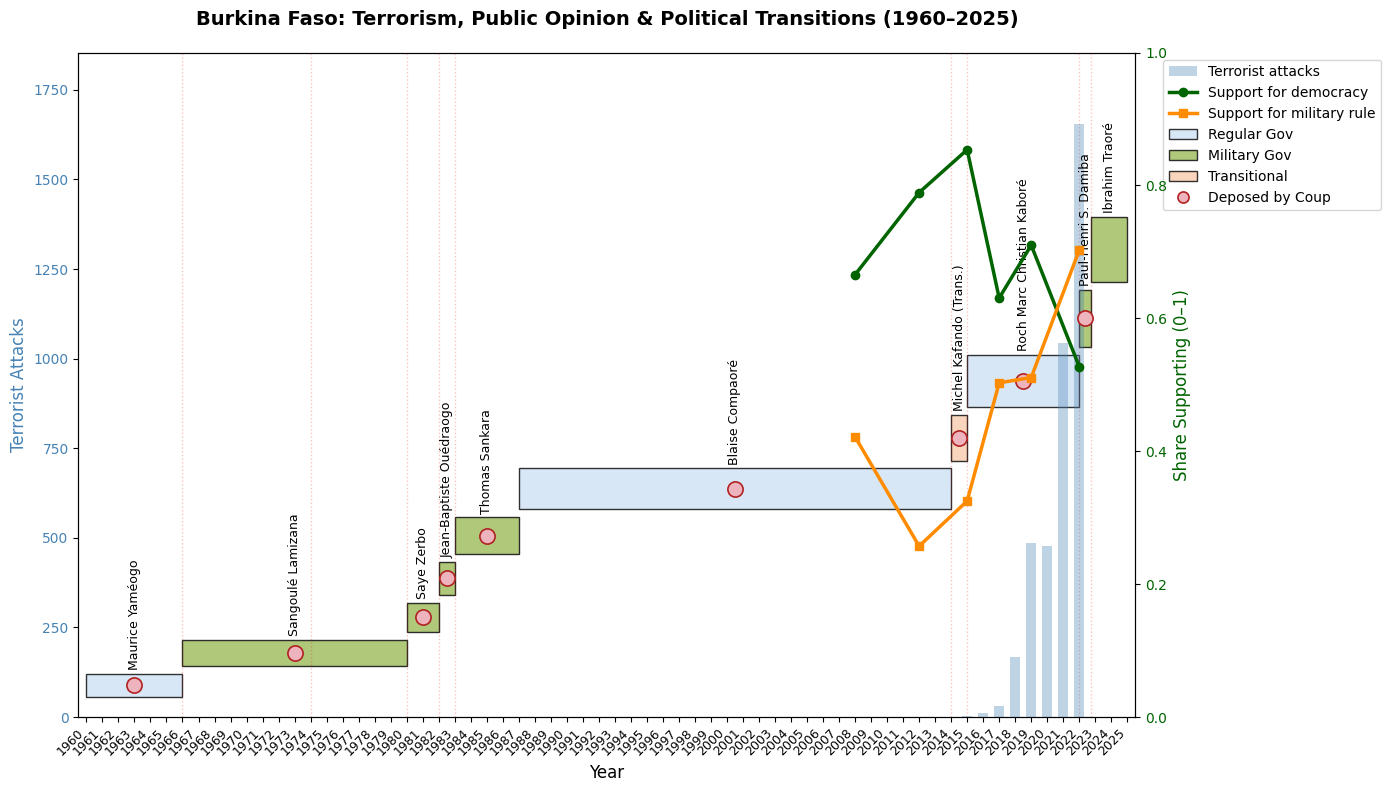}
    \caption{Timeline of Presidential Leadership and Government Types in Burkina Faso (1960-2025)}
    \label{fig1_appendix}
\end{figure}

In Figure \ref{fig1_appendix}, we present a timeline of Burkina Faso's presidents from independence to the present, highlighting the type of government in place and indicating those who were deposed by a coup d’état. The figure also illustrates the evolution of citizens' support for democracy and military rule, alongside the trend in the number of terrorist attacks over time.

Burkina Faso gained independence in 1960, and like many African nations, it experienced early democracy under a one-party state. The first presidential election took place in 1965, and although it was conducted through direct suffrage, it lacked actual democratic competition. The elected president, Maurice Yaméogo, maintained a one-party system that suppressed opposition (\parencite{Ouedraogo2024b}).

Later, the country faced chronic instability, marked by a series of military regimes. It was only with the "third wave" of democratization in the 1990s that Burkina Faso officially adopted a multi-party system. Most African countries adopted this movement and started challenging the long-standing single-party regimes and military governments. However, note that this democratic progress was uneven and fragile. After adopting a multi-party system, Blaise Compaoré was elected president in 1991 and maintained power through authoritarian rule until 2014 \parencite{hilgers_evolution_nodate}. Since 2015, the country's political horizon has been characterized by alternating periods of authoritarian rule.

A few months after Compaoré was ousted, Burkina Faso faced a resurgence of terrorist attacks orchestrated by groups affiliated with Al-Qaeda and the Islamic State. The first attacks, claimed by Ansar Dine, quickly escalated with the emergence of Ansarul Islam in 2016. The consequences of this jihadist insurgency have been profound. In 2022, the Group for the Support of Islam and Muslims (JNIM) carried out over 400 attacks across 10 of Burkina Faso's 13 regions, according to the U.S. Department of State \parencite{USDepartmentofState2023}. The growing insecurity has triggered a severe humanitarian crisis, with the number of internally displaced persons surpassing 2 million—approximately 10\% of the country's population—by early 2024. Furthermore, the 2024 Global Terrorism Index (\parencite{InstituteForEconomicsAndPeace2024}) ranks Burkina Faso as the most terrorism-affected country in the world, surpassing Afghanistan for the first time.

The escalation of terrorist violence has coincided with a sharp deterioration in political stability. Between January 2022 and September 2022, Burkina Faso experienced two military coups that removed Presidents Roch Marc Christian Kaboré and Paul-Henri Sandaogo Damiba from power. In both cases, the military justified its intervention by citing the government's inability to contain the deteriorating security situation. As expressed during the second coup of 2022: "Far from liberating the territories that were occupied, new areas have fallen under the control of armed terrorist groups. Our valiant people have suffered enough and are still suffering" \parencite{Diaw2022}. In addition, public support for these coups reflects the lack of trust in democratic governance \parencite{zie2023mobilisations}. This is shown in \ref{fig1_appendix}, which displays a decline in democratic support from 66\% in 2012 to 52\% in 2022, while support for military rule increased (from 42\% in 2012 to 70\% in 2022).

This positive correlation between political instability and frequent terrorist attacks has created a governance system where authoritarian rule is seen as a natural response to crises in Burkina Faso. The country historical context is characterized by recurrent coups and long periods of military rule, which have weakened the consolidation of democratic norms. Our study aim at showing that terrorism can intensifies this pre-existing tendency.

\section{Conceptual Framework}

In this section, we explain how terrorism can affect political preferences. We propose three main channels through which citizens may prefer military governments.

A primary channel through which terrorism shapes political preferences is the perceived trade-off between security and freedom. Terrorist attacks heighten fear and distress among citizens, leading them to prioritize safety over civil liberties and to support leaders who promise order and stability. In this sense, fear and distress mediate political responses to terrorism by increasing individuals' perception of threat \parencite{doi:10.1111/1467-9280.01433, huq2018terrorism}. Evidence for this mechanism is also found in \parencite{berrebi_are_2008}, \parencite{huq2018terrorism}, and \parencite{AKAY2020103394}. 

Second, terrorism affects trust in democratic institutions by exposing their inability to
respond effectively to crises. This mechanism is what \parencite{AKAY2020103394} called "the rational evaluation". Terrorism can expose a government to scrutiny. In Burkina Faso, as terrorism continues to increase, we observe an erosion of people's trust in civilian governments, which they perceive as incapable of addressing urgent needs. In the case of Burkina Faso, repeated coups in 2022 were justified by military leaders as necessary to address the growing terrorist threat. The deterioration of the security situation was one of the main reasons invoked for the ousting of President Roch Marc Christian Kaboré. As declared in the speech following the second coup of 2022: "Far from liberating the territories that were occupied, new areas have fallen under the control of armed terrorist groups. Our valiant people have suffered enough and are still suffering" \parencite{Diaw2022}. Public support for these coups reflects the lack of trust in democratic governance \parencite{zie2023mobilisations}. Repeated terrorist attacks highlight the perceived inability of
civilian governments to protect citizens, eroding trust in democratic institutions. 

Lastly, terrorism causes population displacement, which results in severe economic
disruption. Terrorism worsens individuals' living and working conditions and might
thus impact individual migration decisions in the aggregate (\parencite{DreherAxel2011Hatw, belmonte_selection_2019}). In Burkina-Faso, over 2 million people have been forced to flee their homes due to terrorist violence \parencite{UNHCR2023}. Most migration happens from regions near the borders
towards the capital (Ouagadougou). However, the economy heavily depends on agriculture,
which occurs in areas from which individuals fly and, thus, creates economic disruption.
This economic disruption by terrorism directly affects the livelihoods of displaced populations. It weakens state capacity. Consequently, individuals may become more supportive of authoritarian leaders or military interventions that promise order and welfare. This economic-insecurity mechanism is particularly relevant in Burkina Faso, where terrorist attacks have displaced millions and devastated rural economies, reshaping both material conditions and political perceptions.

\section{Data Sources}

\subsection{Afrobarometer Dataset}
In order to measure political preference, we draw on data from the \textcite{Afrobarometer2021}. Afrobarometer is a survey that collects public opinion on democracy, governance, and economic and social issues since 1999. Conducted on a two- to three-year cycle across approximately 35 African countries, the survey uses nationally representative, face-to-face interviews. Burkina Faso participated in the 2008, 2012, 2015, 2017, 2019, and 2022 rounds, with interviews conducted in French, Moore, Dioula, and Fulfulde. In addition to political preference variables, the dataset includes socio-demographic characteristics, economic perceptions, and geo-location. We leverage all the available resources for Burkina Faso.

\paragraph{Outcome variables}: To capture attitudes toward democracy, we construct two variables that reflect both normative preferences and perceptions. The first outcome variable classifies respondents as either Pro-Democracy (democracy is always preferable), Conditional Non-Democracy (non-democratic rule may be preferable in some circumstances), or Indifferent (the type of government does not matter).\footnote{This variable is constructed using the self-reported support
for democracy to the questions: Which of these three statements is
closest to your own opinion? Statement 1: Democracy is preferable to any other kind of
government. Statement 2: In some circumstances, a non-democratic government can be
preferable. Statement 3: For someone like me, it does not matter what kind of government we
have." Missing or invalid responses (Do not know, Refused) are excluded}

We also use respondents’ attitudes toward non-democratic governance forms \footnote{Responses range from strongly disapprove to strongly approve. We recode responses as 1 if the respondent selected approve or strongly approve, and 0 if they selected disapprove or strongly disapprove. Missing or invalid responses (Don’t know, Refused) are excluded.} , including their opposition to military rule. From these questions, we construct binary indicators for support for military rule, coded as 1 if the respondent expresses any level of approval and 0 otherwise.

\subsection{Armed Conflict Location \& Event Data}

Terrorism data are obtained from the \textcite{acled} and the Global Terrorism Database \textcite{gtd2021}. ACLED provides open-access data on political violence and protest events worldwide from 1997 to 2024, while the GTD contains open-source records of terrorist attacks from 1970 to 2021.
Our primary analysis relies on ACLED data because it includes the period 2022–2024, during which Burkina Faso experienced the peak of terrorist activity (see Figure~\ref{fig1_appendix} in Appendix~\ref{appendix:fig}). As a robustness, we replicate the analysis using GTD data.

ACLED does not include a successful terrorism indicator. Therefore, we hand-classify each event as a successful terrorist attack or not following a systematic classification procedure described in Appendix~\ref{appendix:classification}. In short, we use actor identities, event types, sub-event classifications, and descriptive notes to identify jihadist violence targeting civilians or consistent with terrorist tactics. 

The main difference between our classification and the GTD is that we include looting and property destruction as terrorist events. In Burkina Faso, such acts are often part of insurgent strategies aimed at intimidation, territorial control, and destabilization. For this reason, and consistent with the context, we classify looting/property destruction as terrorist attacks. As a robustness check, we replicate the analysis excluding these incidents; the results, reported in Appendix~\ref{appendix:tables}, remain consistent.

For each identified terrorist attack, we code whether the attack was successful (1) or failed (0). The detailed success-coding rules appear in Appendix~\ref{appendix:classification}. Events with unknown outcomes (69 observations) are excluded. Terrorist attacks are then aggregated at the department–year level; a department–year cell is coded as successful if at least one attack in that year was successful.

\subsection{Department Characteristics}
We check for balance across a wide range of characteristics in departments that experienced successful or failed attacks. Data on these characteristics are taken from different sources. The other sources used are presented synthetically in Table \ref{tab:data_sources} in Appendix \ref{appendix:tables}.

\subsection{Geographic matching}
On the one hand, we have the Afrobarometer data at the individual level; on the other hand, we have the event data. To consistently merge these two sources, we use the settlement dataset, which contains all known settlements in Burkina Faso along with their geographic coordinates. For both individual and department-year observations, each record is matched to the settlement database using the region variable (Regions are the first administrative level, while departments are the third administrative level in Burkina Faso). We then compute the minimum distance from each observation to the nearest department centroid, identify the corresponding department, and assign it as the observation's location.

\section{Identification Strategy}
\subsection{Randomness of Successful Terrorist Attacks}
Our analysis assumes that the success of a terrorist attack is exogenous to the department's characteristics. In this section, we test for pre-attack balance across a list of department characteristics. To this end, we use the variable $SUCCESS_s$ (=1 if at least one successful terrorist attack happened in a given department-year ); the variable is missing for departments that did not experience any
attacks. Then, we regress each pre-attack department characteristic (measured in the year just before the attack) $\mathbf{X}_{s, t}$ on the success indicator. The equation we estimate is the following : 
\begin{equation}
\label{eq1}
\mathbf{X}_{s, t = t_{\text{ATTACK}-1}} = \beta_0 + \beta_1 \textit{SUCCESS}_s + \varepsilon_s
\end{equation}

If the coefficient on $SUCCESS_s$ ($\hat{\beta_1}$) is essentially zero, it indicates that the treated and untreated departments were similar before the attack, which supports the validity of our identification strategy. We present the results of this test in Columns 3 and 4 in Table \ref{table1}. Panel A of Table \ref{table1} tests whether department hit by successful attacks differ from those hit by failed attacks before the attack occurs. In Panel B, we are testing successful and failed attacks are not systematically different in their nature. If attack types between successful and failed attacks, then success might not be random, and we must account for these differences in our analysis.

The results show that successful attacks are not related to key socioeconomic factors that could bias our analysis. For example, direct access to specific natural resources influences the likelihood of an attack's success. As shown in Table  \ref{table1}, however, success is uncorrelated to access to natural resources. 

Another potential concern relates to the level of criminality. Criminal environments create more opportunities for terrorist groups to blend in and move around without attracting attention. The results in Table 1, however, make clear that success does not exhibit
such selection. Finally, we find that success is uncorrelated with the department-level education, tourism, health, economics, and demographics. 

As expected, the distance from Kidal (Mali) and Tilabéri (Niger) is a predictor of the success of an attack. Indeed, these cities are known hotspots for armed groups, making the surrounding areas more vulnerable to attacks. Similarly, the level of institutional service is a strong predictor of the likelihood of an attack's success. Institutional service can matter because departments with more substantial administrative capacity have better state presence and coordination, making it harder for terrorists to carry out a successful attack. We include these variables in our analysis. In panel B, we compare observable characteristics of all the attacks in our sample. As expected, the number of attempted attacks and the type of attack are statistically significant. We include these variables in our analysis because the more attempts there are, the higher the chance that at least one attack will succeed, and attack types are predictive of success. 
{\footnotesize 
\renewcommand{\arraystretch}{0.9} 
\renewcommand{\tablename}{Table}

\begin{longtable}{l*{5}{r}}

\caption{Predict Terror Attack}\label{table1}\\

\toprule
Variable & $\bar{Y_0}$ & $\bar{Y_1}$ & $\hat{\beta}$ &
p-value $H_0:\beta=0$ & N \\
\midrule
\endfirsthead

\multicolumn{6}{l}{\tablename~\thetable{}: Predict Terror Attack \textit{(continued)}}\\
\toprule
Variable & $\bar{Y_0}$ & $\bar{Y_1}$ & $\hat{\beta_1}$ &
p-value $H_0:\beta_1=0$ & N \\
\midrule
\endhead

\midrule
\multicolumn{6}{r}{\textit{(continued on next page)}}\\
\endfoot

\bottomrule
\multicolumn{6}{p{16cm}}{\footnotesize
\textit{Notes:} This table presents estimated regression coefficients (column 3) and corresponding p-values (column 4) estimating equation \ref{eq1}. Panel A compares pre-attack characteristics between department targeted with successful and failed attacks, using values measured in the year immediately preceding the attack. Panel B reports cross-sectional comparisons of attack-level characteristics.
}\\
\endlastfoot

\multicolumn{6}{l}{\textbf{Panel A. Department characteristics}}\\
\midrule
\emph{Demographic:}\vspace*{.2em} \\ \hspace{.5em} Population&10.64&10.62&-0.03&0.88&188\\
 \hspace{.5em} Number of deaths&20.11&12.52&-7.60&0.37&98\\
 \hspace{.5em} Number of births&1914.19&2213.19&299.00&0.55&98\\
\emph{Economic:}\vspace*{.2em} \\ 
\hspace{.5em} Tax revenue per capita&2905.76&1342.87&-1562.89&0.31&186\\
 \hspace{.5em} Cereal production by province&11.76&11.56&-0.19&0.18&181\\
\emph{Geographic \& Resources:}\vspace*{.2em} \\ \hspace{.5em} Urban&0.12&0.14&0.01&0.87&188\\
 \hspace{.5em} Distance to Kidal (km)&850.24&748.06&-102.19&0.00&188\\
 \hspace{.5em} Distance to Ouagadougou (Capital) (km)&189.46&221.53&32.07&0.18&188\\
 \hspace{.5em} Distance to Tillabéri (km)&532.59&415.94&-116.65&0.00&188\\
 \hspace{.5em} Distance to International Airport (km)&190.93&222.04&31.11&0.19&188\\
 \hspace{.5em} Distance to water (km)&69.51&64.89&-4.63&0.69&188\\
 \hspace{.5em} Distance to onshore Petroleum (km)&1062.28&1044.90&-17.38&0.60&188\\
 \hspace{.5em} Distance to lootable gold deposit (km)&136.23&134.24&-1.99&0.91&188\\
 \hspace{.5em} Distance to gemstone deposit (km)&613.45&665.10&51.65&0.07&188\\
 \hspace{.5em} Distance to road (km)&11752.06&11503.81&-248.24&0.80&188\\
\emph{Public services \& institutions:}\vspace*{.2em} \\ 
\hspace{.5em} Public services (points)&68.40&68.00&-0.40&0.88&186\\
 \hspace{.5em} Institutional capacity (Points)&49.78&42.57&-7.20&0.09&186\\
\emph{Education:}\vspace*{.2em} \\ \hspace{.5em} Primary school enrollment&659.17&510.23&-148.94&0.32&187\\
 \hspace{.5em} Enrollment in Primary School&8030.62&6914.86&-1115.76&0.51&186\\
\emph{Road accidents \& crime:}\vspace*{.2em} \\ \hspace{.5em} Traffic accidents&1757.75&1076.66&-681.09&0.18&187\\
 \hspace{.5em} Number of scams&118.56&71.14&-47.42&0.08&187\\
 \hspace{.5em} Number of traffic accidents&1586.81&734.94&-851.87&0.12&187\\
\emph{Health:}\vspace*{.2em} \\ \hspace{.5em} Number of infants 0-11 vaccinated&900.29&1286.53&386.24&0.21&187\\
 \hspace{.5em} Number of assisted deliveries&2589.33&2075.94&-513.40&0.72&112\\
\emph{Tourism / housing:}\vspace*{.2em} \\ \hspace{.5em} Number of rooms&328.81&269.24&-59.58&0.64&173\\
\multicolumn{6}{l}{\textbf{Panel B. Attack characteristics}}\\
\emph{Attack characteristics::}\vspace*{.2em} \\ 
 \hspace{.5em}  Number of attacks&1.00&7.71&6.71&0.00&505\\
 \hspace{.5em} Armed Assault Attacks&0.55&0.98&0.43&0.00&530\\
 \hspace{.5em} Bombing/Explosion Attacks&0.09&0.67&0.58&0.00&530\\
 \hspace{.5em} Facility/Infrastructure Attacks&0.36&0.92&0.55&0.00&530\\
\end{longtable}
}

\subsection{The Model}
We use a stacked Difference-in-Differences approach to estimate how successful terrorist attacks influence political preferences \parencite{10.1093/qje/qjz014, 10.1257/pol.20180076, https://doi.org/10.1111/ecin.12027,10.1257/aer.20201524}. This method is well-suited to our context because political attitudes are observed only in specific years (2008, 2012, 2015, 2017, 2019, and 2022), and we have a staggered treatment with entry and exit, since successful attacks can occur at different times, and departments can move in and out of treatment.
 
The stacked DID estimator reorganizes the data into sub-experiments based on treatment timing. Each sub-experiment focuses on a specific treatment year $a$, aligning observations by event time relative to treatment rather than calendar time. The treatment group consists of all units first treated in year $a$. The control group includes all units that have not yet received treatment by year $a$, which include both never-treated units and those that will be treated in later periods. Observations are restricted to an event window covering $\kappa_{pre}$  years before and $\kappa_{post}$ years after treatment. Units that are re-treated within the event window are excluded.

The resulting sub-experiments are then stacked into a single dataset, ensuring consistent comparisons across time. Using the final stacked dataset, we estimate the following model : 

\begin{equation}
    \label{eq2}
    Y_{isae} = \alpha 
    +  \sum_{\substack{e=-3 \\ e \neq 0}}^{2} 
    \beta_e \left( D_{sae} \times \mathbf{1}[e] \right)
    +  X'_{isae} \Gamma 
    + m_{sa}
    + v_{ae}
    + \epsilon_{isae},
\end{equation}

where \( Y_{isae} \) denotes the outcome variable for individual \( i \) in unit (department) \( s \), in sub-experiment  \( a \), and in event time \( e \); \( D_{sae} \) is a treatment indicator; \( \mathbf{1}[e] \) represents event-time dummies; \( X'_{isae}  \Gamma \) is the vector of individual-level control variables; \( m_{sa} \) denotes $units \times sub-experiment$  fixed effects;  \( v_{ae} \) denotes $event-time \times sub-experiment$  fixed effects; and \(\epsilon_{isae} \) is the error term. 

The stacked event study is estimated using corrected sample weights proposed by \textcite{NBERw32054}. We use $\kappa_{\text{pre}} = 3$ and $\kappa_{\text{post}} = 2$. Event time $e=-1$ is the omitted (reference) period. We also report the results for different windows in the Appendix \ref{appendix:tables}. In addition to the event-study estimate, we report the aggregated average treatment effect on the treated\footnote{. 
The average post-treatment effect over the first 3 post years is : 
\begin{equation}
\hat{\theta}^{\text{post}}_{\kappa} = \frac{1}{3} \left( \hat{\beta}_{e=0} + \hat{\beta}_{e=1} + \hat{\beta}_{e=2} \right)
\end{equation}
}

\section{Results}
We begin this section by presenting the estimated effects of successful terror attacks on political attitudes. It then proceeds with a series of checks to validate the identification assumptions and assess the robustness of the findings. 

\subsection{Main Findings}
\subsubsection{Dynamic Effects}

\begin{figure}[ht]
    \centering
    \begin{subfigure}{0.48\textwidth}
        \centering
        \includegraphics[width=\textwidth]{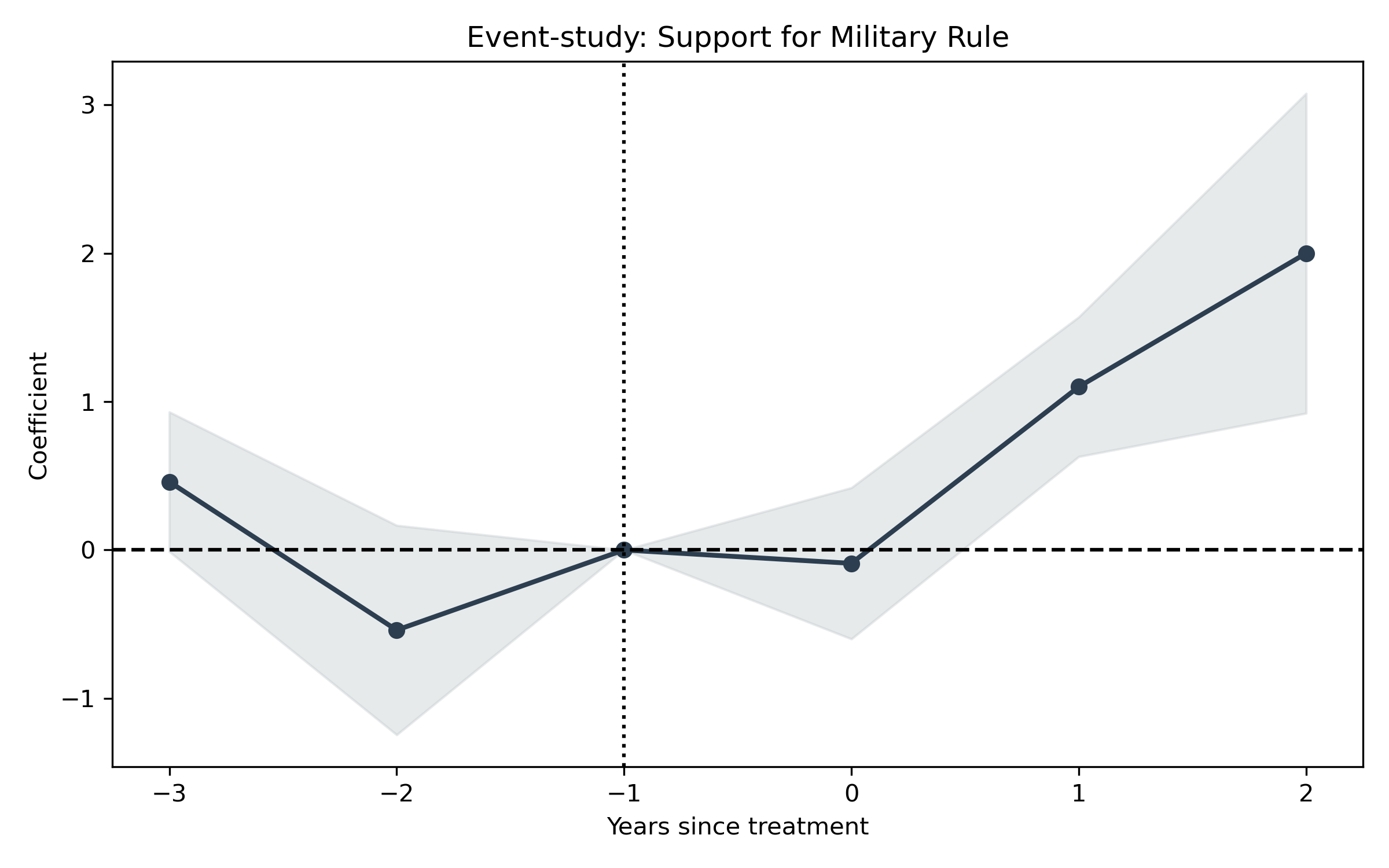}
        \caption{Support for Military Rule}
        \label{fig:militaryrule}
    \end{subfigure}
    \hfill
    \begin{subfigure}{0.48\textwidth}
        \centering
        \includegraphics[width=\textwidth]{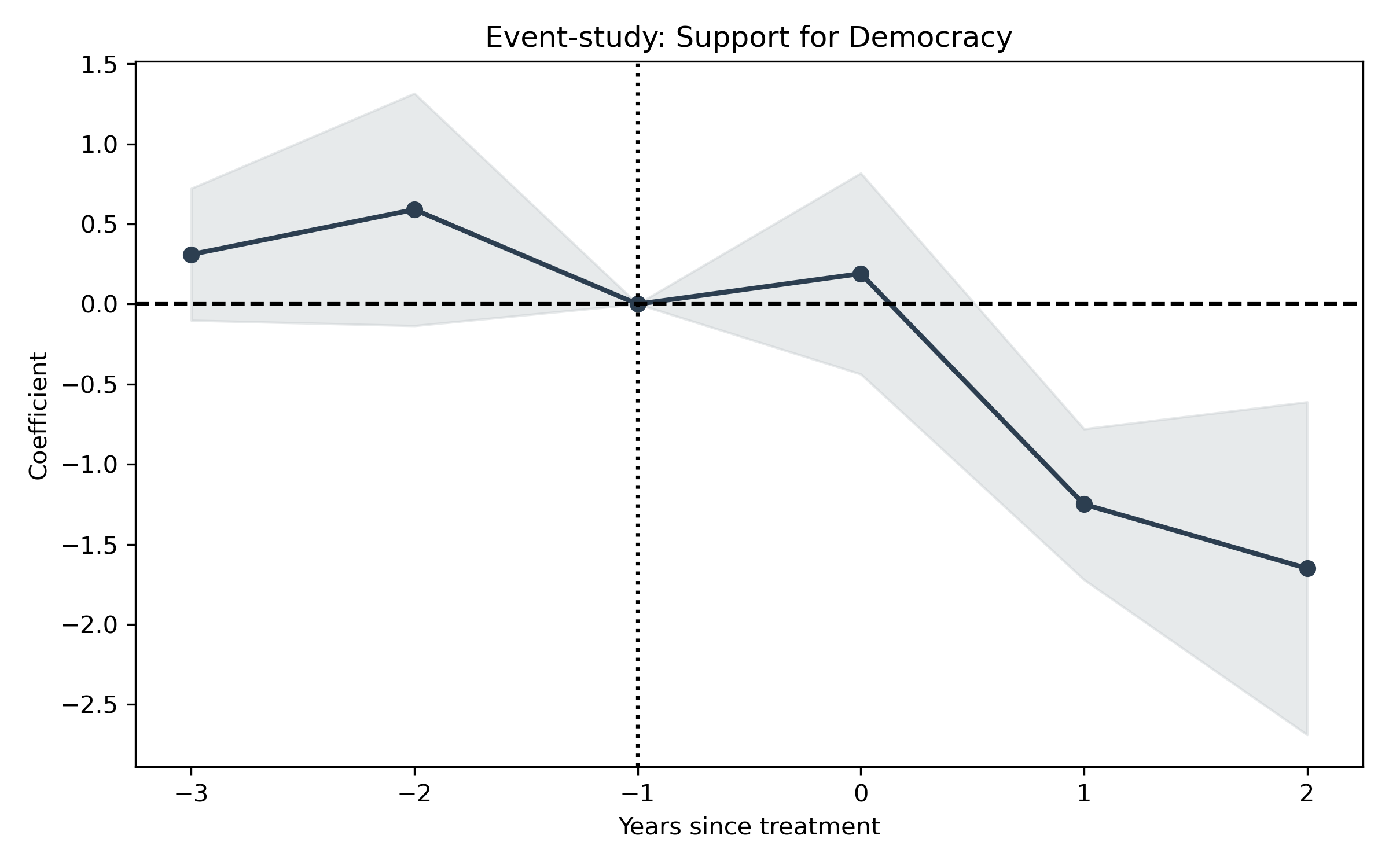}
        \caption{Support for Democracy}
        \label{fig:democracy}
    \end{subfigure}
    \caption{Dynamic Treatment Effects of Terrorist Attacks on Political Preferences}
    \label{fig:eventstudy_combined}
\end{figure}

In Figure \ref{fig:eventstudy_combined}, we plot the dynamic effects of exposure to terrorist attacks on political preferences estimated from equation \ref{eq2}, with 95\% confidence intervals based on standard errors clustered at the department level. 

For both outcomes, the pre-treatment coefficients oscillate around zero and are statistically insignificant, which supports the validity of the parallel-trends assumption. The effects of terrorism do not appear immediately: statistically significant changes in support for military rule and support for democracy emerge only one year after the attack, and these effects persist into the second post-treatment year. This pattern suggests that the political consequences of terrorism are both delayed and persistent.

While the event-study plot provides a visual check of parallel trends, pre-trend tests may have limited information about the parallel-trends assumption. To address possible violations, we implement the sensitivity analysis proposed by \textcite{RambachanAshesh2023AMCA}. \textcite{RambachanAshesh2023AMCA} provides robust inference under controlled departures from parallel trends. 

We focus on the first and second post-treatment coefficients and evaluate their robustness under relative magnitude and smoothness restrictions. 
The relative-magnitude approach examines whether any post-treatment deviation from parallel trends exceeds the most significant deviation observed before treatment. If post-treatment deviations substantially exceed this benchmark, the estimates may be unreliable. The smoothness restriction instead tests whether post-treatment patterns could reflect a continuation of the pre-treatment trend; deviations from such smooth extrapolation may indicate bias. For each value of $M$, the procedure computes robust confidence intervals that remain valid under deviations up to that magnitude. 

\begin{figure}[ht]
    \centering
    \begin{subfigure}[t]{0.48\textwidth}
        \centering
        \includegraphics[width=\linewidth]{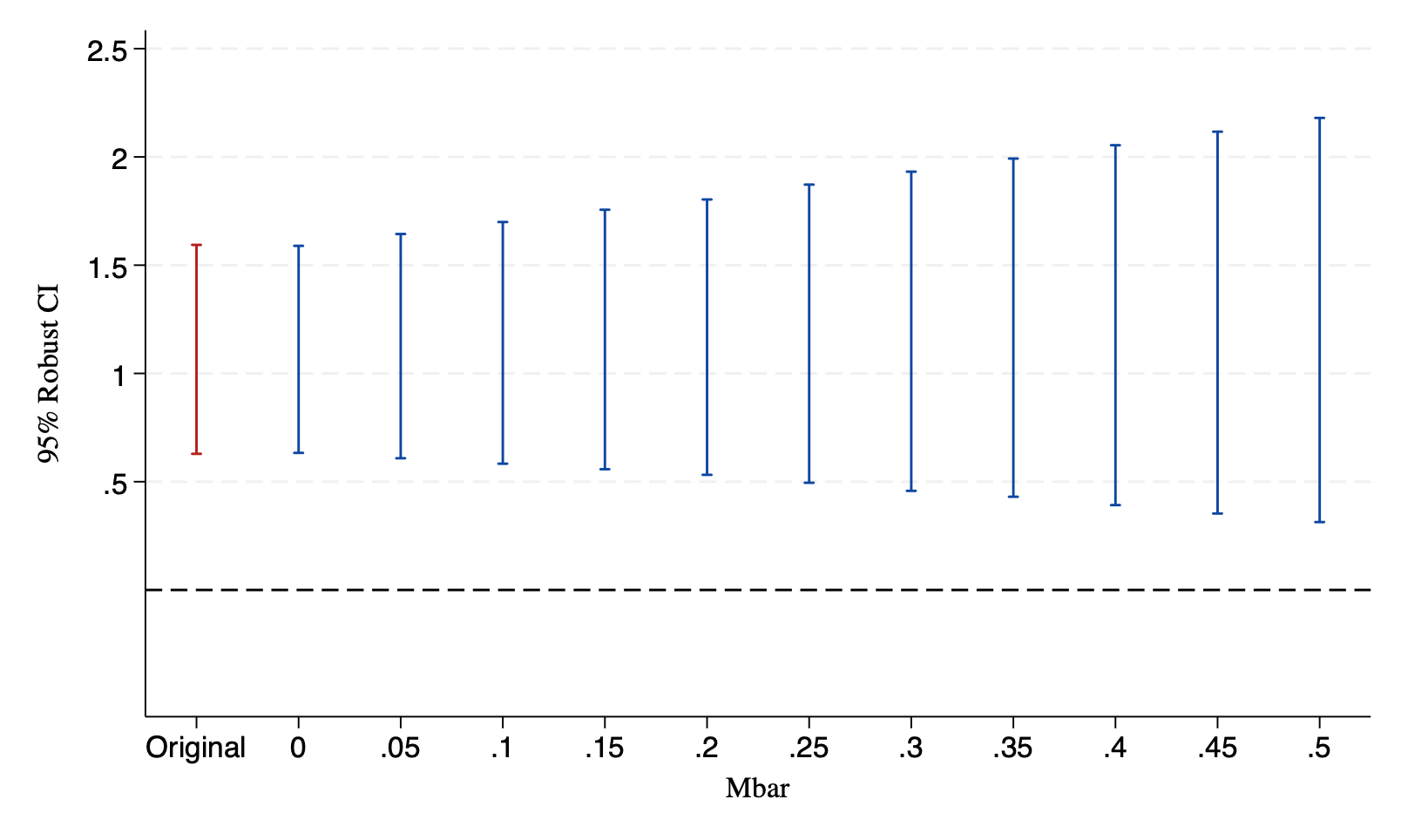}
        \caption{Relative magnitude – Support for military rule}
        \label{fig:military_relative}
    \end{subfigure}
    \hfill
    \begin{subfigure}[t]{0.48\textwidth}
        \centering
        \includegraphics[width=\linewidth]{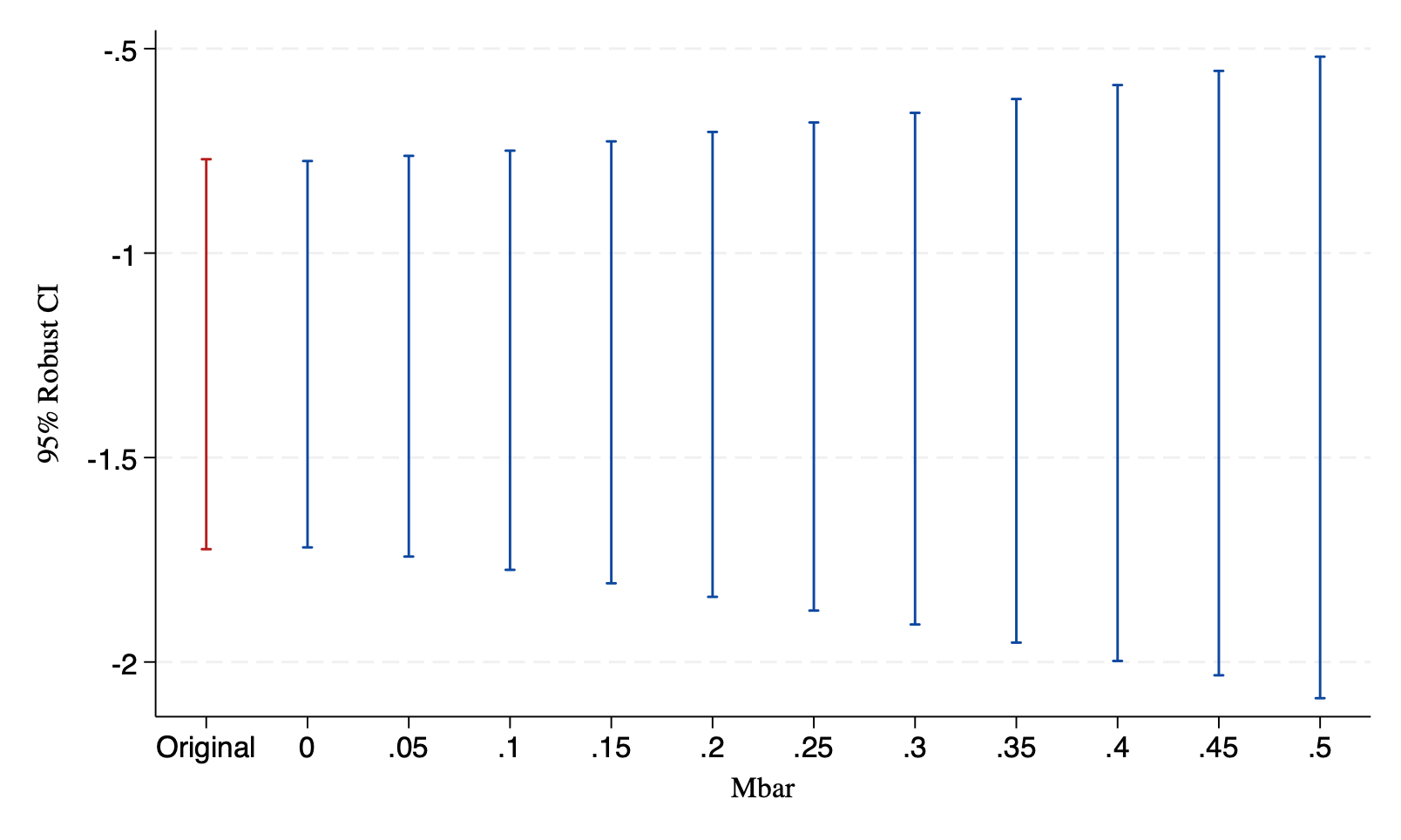}
        \caption{Relative magnitude – Support for democracy}
        \label{fig:dem_relative}
    \end{subfigure}

    \vspace{0.7em}

    \begin{subfigure}[t]{0.48\textwidth}
        \centering
        \includegraphics[width=\linewidth]{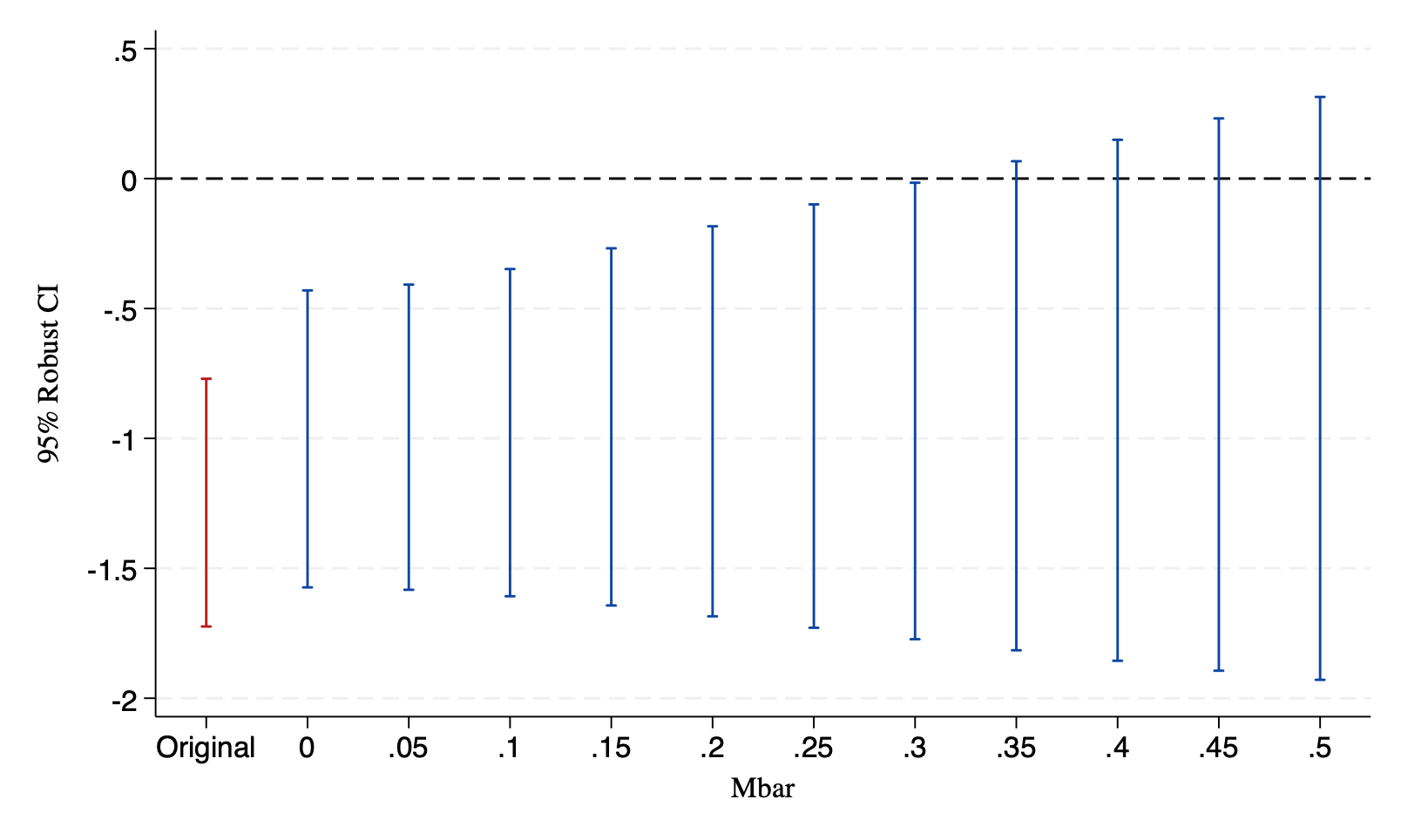}
        \caption{Smoothness – Support for military rule}
        \label{fig:military_smooth}
    \end{subfigure}
    \hfill
    \begin{subfigure}[t]{0.48\textwidth}
        \centering
        \includegraphics[width=\linewidth]{Graph_and_Tables/LatexGraphs/pref3_honestdid_smoothness.png}
        \caption{Smoothness – Support for democracy}
        \label{fig:dem_smooth}
    \end{subfigure}

    \caption{Sensitivity of estimated treatment effects to violations of parallel trends,
    using the relative magnitude (top row) and smoothness (bottom row) restrictions.}
    \label{fig:honestdid_4panel}
\end{figure}

Figure \ref{fig:honestdid_4panel} plots the 95\% confidence intervals for the year-1 and year-2 treatment effects. The red bars show conventional estimates, while the blue bars display confidence intervals adjusted for potential violations of parallel trends for increasing values of M. Under the relative-magnitude restriction (Panel A), the estimated effects remain statistically significant even when allowing for sizable departures from parallel trends. Under the smoothness restriction, the effects remain significant for deviations up to roughly M = 0.35, indicating robustness to moderate departures from linear pre-trend extrapolation.
Overall, the sensitivity analysis shows that the year-1 and year-2 treatment effects are robust to modest violations of the parallel-trends assumption.

\subsubsection{Baseline Treatment Effect}
Table \ref{table2a} presents the aggregated average treatment effect using the coefficients from equation \ref{eq2}. 

Columns (1) and (2) report statistically significant impacts for both variables. A coefficient of 1.00 means that after a successful terror attack, the probability that a respondent approves of military rule increases by about 1 point. Similarly, after a successful terrorist attack occurs in a department, the likelihood of reporting support for a democratic government decreases by 0.91 points.
\begin{table}[ht!]
\centering
\begin{threeparttable}
\renewcommand{\tablename}{Table}

\caption{Successful Terror Attacks and Political Attitudes}
\label{table2a}

\begin{tabular}{lcc}
\hline\hline
            & (1) & (2) \\
            & Military rule & Support democracy \\
\hline
Treated $\times$ Post & 1.00$^{***}$ & -0.91$^{**}$ \\
                      & (0.29)       & (0.29)       \\
\hline
R-Squared             & 0.15         & 0.14         \\
N                     & 11,098       & 10,979       \\
\hline
Sub-experiment $\times$ Departments FEs & Yes & Yes \\
Sub-experiment $\times$ Event-time FEs  & Yes & Yes \\
Attacks types FEs                       & Yes & Yes \\
Individual controls                     & Yes & Yes \\
Departments covariates                  & Yes & Yes \\
\hline\hline
\end{tabular}

\begin{tablenotes}
  \footnotesize
  \item Standard errors in parentheses are clustered at the department level. $^{*} p<0.05$, $^{**} p<0.01$, $^{***} p<0.001$.
\end{tablenotes}

\end{threeparttable}
\end{table}

\subsubsection{Examining potential spillover effects}

A crucial assumption for the validity of our stacked DiD design is the Stable Unit Treatment Value Assumption (SUTVA), which requires that the treatment in one unit does not affect outcomes in untreated units. In our context, SUTVA would be violated if a successful terrorist attack in department $d$ has an effect on political preferences among individuals living in neighboring departments.

To address this concern, we re-estimate Equation \ref{eq2} while additionally controlling for the occurrence of successful attacks in neighboring departments,\footnote{Neighbors are defined using border adjacency.} thereby accounting for potential spillover effects across administrative boundaries. The results are presented in Table \ref{table2i}.

Table \ref{table2i} reports the dynamic effects and average treatment of successful terrorist attacks on political attitudes after controlling for attacks in neighboring departments. Across both outcomes, support for military rule (Column 1) and support for democracy (Column 2), the inclusion of neighboring-attacks controls leaves the main pattern of results essentially unchanged. Immediate effects remain small and statistically insignificant, consistent with the baseline event-study results. Effects start showing 1 year after treatment, and appear persistent over time.

\begin{table}[ht!]
\centering
\begin{threeparttable}
\renewcommand{\tablename}{Table}

\caption{Successful Terror Attacks and Political Attitudes with Spillover}
\label{table2i}

\begin{tabular}{lcc}
\hline\hline
            & (1) & (2) \\
            & Military rule & Support democracy \\
\hline            
Treated $\times$ Event-time, -3&        0.10         &        0.21         \\
            &      (0.22)         &      (0.17)         \\
[1em]
Treated $\times$ Event-time, -2&       $-0.78^{**}$ &        0.32         \\
            &      (0.29)         &      (0.33)         \\
[1em]
Treated $\times$  Event-time, 0&       -0.40         &       -0.07         \\
            &      (0.25)         &      (0.29)         \\
[1em]
Treated $\times$  Event-time, +1&        $1.70^{***}$&       $-1.46^{***}$\\
            &      (0.33)         &      (0.29)         \\
[1em]
Treated $\times$  Event-time, +2&        $3.27^{***}$&       $-1.87^{**}$ \\
            &      (0.64)         &      (0.61)         \\
\hline
Treated $\times$ Post&$1.52^{***}  $       &$-1.13^{**} $        \\
            &       ( 0.35)         &        (0.34 )        \\

\hline
R-Squared   &        0.15         &        0.14         \\
N           &    11,098    &    10,979        \\

\hline
Sub-experiment $\times$ Departments FEs & Yes & Yes \\
Sub-experiment $\times$ Event-time FEs  & Yes & Yes \\
Attacks types FEs                       & Yes & Yes \\
Individual controls                     & Yes & Yes \\
Departments covariates                  & Yes & Yes \\
\hline\hline
\end{tabular}

\begin{tablenotes}
  \footnotesize
  \item Standard errors in parentheses are clustered at the department level. $^{*} p<0.05$, $^{**} p<0.01$, $^{***} p<0.001$.
\end{tablenotes}

\end{threeparttable}
\end{table}

For support for military rule, the event-time coefficient at $ e - 2$ is statistically significant. While this indicates a small deviation from perfect parallel trends, there is no consistent pattern of pre-treatment divergence, and the remaining pre-period coefficients are close to zero. Moreover, the robustness checks using the \textcite{RambachanAshesh2023AMCA} framework confirm that the post-treatment effects remain stable under moderate violations of the parallel-trends assumption, mitigating concerns raised by the isolated pre-trend significance.

\subsection{Robustness checks}
We conduct several robustness checks to assess the credibility of the estimated causal effects of successful attacks on political preferences.

First, we tighten the definition of our control group by considering only the treated department. As reported in Table \ref{table2j}, the results remain consistent. In addition, for support for military rule, the event-time coefficient at $ e=-3$ appears now statistically significant.

In repeated cross-sectional DiD analyses, changes in the composition of treatment and control groups over time can violate the parallel trends assumption, leading to biased estimates \textcite{SantAnnaPedroHC2023DwCC}. We employ the Entropy Balancing Matching (EBM) method proposed by \textcite{HainmuellerJens2012EBfC} to enhance the comparability of individuals in treatment and control groups, which eases the potential compositional change. The EBM focuses on balancing the moment conditions of given variables \footnote{The variables we consider are: age and education, religion, lived poverty index, and employment}. Results are presented in \ref{table4a} and are consistent.

Second, to address potential concerns about our event classification, we replicate the analysis using the \textcite{gtd2021} dataset. To ensure comparability with the main specification, we apply the Entropy Balancing Matching (EBM) method at the individual level. However, due to collinearity arising from missing data, not all control variables from the baseline model could be included in this specification. Consequently, while the GTD-based results broadly support our baseline findings, they should be interpreted with caution. Results are presented in \ref{table2c} and are consistent. 

Using the \textcite{gtd2021} dataset, we find an immediate and statistically significant impact of successful terrorist attacks on both political attitudes. In the year of the attack (event time 0), support for military rule increases significantly, while support for democracy decreases. One year after the attack, the effect on military support reverses, becoming significantly negative, while support for democracy slightly rebounds, though not significantly. Two years after the attack, we again observe a significant rise in support for military rule, accompanied by a renewed decline in democratic support. Overall, the average treatment effect remains statistically significant for both outcomes, although the magnitude of the coefficients is relatively modest. In addition, we replicate the analysis without including Looting as a terrorist attack (see Table \ref{table2b}). In this analysis, the effect does not persist in the second year.

These differences in trends over time may reflect variations in the composition of treated departments across datasets. As shown in Table~\ref{tab:treated_only}, the treated groups differ between the GTD and ACLED samples.

Second, to ensure the robustness of our results and rule out the influence of spurious correlations with unobserved factors, we perform a placebo test. We randomly reassign the year of first exposure to a successful terrorist attack across departments while keeping the original distribution of adoption years. This procedure breaks the true relationship between treatment timing and outcomes, allowing us to generate a distribution of placebo estimates under the null hypothesis of no effect. 
For each of 700 random permutations, we rebuild the stacked dataset by treating the permuted adoption years as if they were the actual treatment dates. We then replicate the exact same estimation procedure as in the baseline analysis, including the exact event-study specification, fixed effects, controls, and weighting scheme, to obtain a single average treatment effect (ATT) from each iteration. The collection of these 700 placebo ATTs forms an empirical distribution centered around 0 (see Figure \ref{fig:placebo_combined} in the Appendix \ref{appendix:fig})

Lastly, we re-estimate the model using alternative pre- and post-treatment windows. In addition to the baseline specification with a window of $[3,2]$, we consider longer or more balanced horizons, specifically $[4,2]$, $[5,2]$, $[3,3]$, and $[4,3]$. Across all specifications, the results reveal a consistent short-run reaction to terrorist attacks. Support for military rule rises sharply and significantly one year after the attack, indicating that citizens temporarily shift toward favoring authoritarian responses when security is threatened. However, this effect fades or reverses within one to two years. In contrast, support for democracy declines slightly one year after the attack, though the decrease is smaller and often not statistically significant. The decrease persists in the second year.

Taken together, these findings point to a short-term trade-off between security and democratic ideals: terrorism triggers a brief surge in preference for military governance, while democratic support remains relatively resilient over time.

\section{Mechanisms}
\subsection{Trade off between security \& freedom}

To examine one of the mechanisms underlying these results, we explore whether terrorist attacks affect individuals' preferences in the trade-off between security and freedom. The Afrobarometer survey provides a direct measure of this attitude in Rounds 7 and 8\footnote{Respondents were asked to choose between two statements: Statement 1 — "Even if faced with threats to public security, people should be free to move about the country at any time of day or night," and Statement 2 — "When faced with threats to public security, the government should be able to impose curfews and set up special roadblocks to prevent people from moving around." }. Based on this question, we construct two binary indicators: one capturing whether an individual prioritizes freedom over security, and another capturing whether they prioritize security over freedom. We then estimate the average treatment effect of exposure to a terrorist attack on these two indicators to assess whether terrorism shifts citizens' preferences toward valuing security at the expense of freedom.

\begin{table}[ht!]
\centering
\begin{threeparttable}
\renewcommand{\tablename}{Table}

\caption{Trade-off between Security \& Freedom}
\label{table3c}

\begin{tabular}{lccc}
\hline\hline
            & (1) & (2) & (3) \\
            & Freedom over security & Security over freedom &Fear\\
\hline
Treated $\times$ Post&       $-1.45^{***}$&        $1.45^{***}$&        $1.22^{*}$  \\
            &      (0.37)         &      (0.37)         &      (0.62)         \\
\hline
R-squared   &        0.15         &        0.15         &        0.19         \\
N           &     6,991        &     6,991        &     6,991         \\
\hline
Sub-experiment $\times$ Departments FEs & Yes & Yes& Yes \\
Sub-experiment $\times$ Event-time FEs  & Yes & Yes& Yes \\
Attacks types FEs                       & Yes & Yes& Yes \\
Individual controls                     & Yes & Yes& Yes \\
Departments covariates                  & Yes & Yes& Yes \\
\hline\hline
\end{tabular}

\begin{tablenotes}
  \footnotesize
  \item Standard errors in parentheses are clustered at the department level. $^{*} p<0.05$, $^{**} p<0.01$, $^{***} p<0.001$.
\end{tablenotes}

\end{threeparttable}
\end{table}

Table \ref{table3c} examines the effect of terrorist attacks on individuals' preferences regarding the trade-off between security and freedom. The first two columns present estimates for respondents prioritizing freedom over security and security over freedom, respectively, while Column (3) investigates whether these shifts are associated with a rise in fear of extremist violence. All specifications include department-by-sub-experiment and sub-experiment-by-event-time fixed effects, as well as controls for attack types, individual characteristics, and departmental covariates.

The results show that exposure to a terrorist attack significantly reduces the likelihood of prioritizing freedom by about 1.45 points and correspondingly increases the likelihood of prioritizing security by the same magnitude. This mirrored pattern suggests a clear trade-off between freedom and security preferences following terrorist incidents. Column (3) further supports the proposed mechanism: exposure to terrorism increases the likelihood of fearing violence by extremist groups by approximately 1.22 points. Together, these findings indicate that terrorist attacks trigger a rise in fear and perceived insecurity, which in turn drives individuals to favor stronger security measures, even at the cost of certain personal freedoms.

\subsection{Trust Erosion in democratic institutions}
We now turn to a second potential mechanism linking terrorism and citizens' perceptions of government competence. Terrorist attacks may not only generate fear but also reveal the state's inability to guarantee public safety. In countries where democratic institutions remain fragile or underdeveloped, such exposure can intensify doubts about the government's capacity to protect its citizens, undermining trust in both state institutions and democracy more broadly.

To examine this hypothesis, we use data from Afrobarometer that capture respondents' trust in key political and institutional actors, including the President, Parliament, Electoral Commission, and the military. 
\begin{table}[ht!]
\centering
\begin{threeparttable}
\renewcommand{\tablename}{Table}

\caption{Trust in Democratic Institutions}
\label{table3a}

\begin{tabular}{lccc}
\hline\hline
            & (1) & (2) & (3) \\
&{President trust}&{Parliament trust}&{Army trust} \\
\hline
Treated $\times$ Event-time, -3&        0.20         &       -0.31         &        0.25         \\
            &      (0.23)         &      (0.21)         &      (0.18)         \\
[1em]
Treated $\times$ Event-time, -2&        0.26         &       $-1.08^{**}$ &        0.10         \\
            &      (0.30)         &      (0.34)         &      (0.28)         \\
[1em]
Treated $\times$  Event-time, 0&        0.30         &       $-0.77^{**}$ &        0.15         \\
            &      (0.28)         &      (0.26)         &      (0.22)         \\
[1em]
Treated $\times$  Event-time, +1&      $ -0.91^{***}$&       -0.22         &      $ -0.86^{***}$\\
            &      (0.19)         &      (0.24)         &      (0.23)         \\
[1em]
Treated $\times$  Event-time, +2&       $-1.29^{**}$ &       -0.16         &      $ -1.03^{*}$  \\
            &      (0.39)         &      (0.47)         &      (0.46)         \\
\hline
Treated $\times$ Post&$-0.64^{**} $        & $-0.38^{}$         &$-0.58^{*} $        \\
            &        (0.23 )        &       ( 0.26)         &        (0.23 )        \\
R-Squared   &        0.16         &        0.17         &        0.13         \\
N      &    11,068     &    11,186      &    11,120     \\
\hline
Sub-experiment $\times$ Departments FEs & Yes & Yes& Yes \\
Sub-experiment $\times$ Event-time FEs  & Yes & Yes& Yes \\
Attacks types FEs                       & Yes & Yes& Yes \\
Individual controls                     & Yes & Yes& Yes \\
Departments covariates                  & Yes & Yes& Yes \\
\hline\hline
\end{tabular}

\begin{tablenotes}
  \footnotesize
  \item Standard errors in parentheses are clustered at the department level. $^{*} p<0.05$, $^{**} p<0.01$, $^{***} p<0.001$.
\end{tablenotes}

\end{threeparttable}
\end{table}

The results in Table~\ref{table3a} reveal distinct patterns across institutions. Trust in the President declines sharply and significantly in the years immediately following an attack (event times $+1$ and $+2$), suggesting that citizens lose confidence in the government’s capacity to maintain order and protect them. By contrast, the estimated effects on trust in Parliament are smaller and less systematic. 

For this outcome, coefficients at event times $-2$ and $0$ are already negative and statistically significant, while post-attack effects are weaker and mostly insignificant. This pre-treatment pattern raises concerns about a possible deviation from parallel trends, indicating that trust in Parliament may have been declining even before the attacks. Moreover, the dynamic pattern appears consistent with a convergence in trust levels between treated and control departments after exposure, rather than a sharp post-attack decline.

Surprisingly, trust in the army declines sharply and significantly in the years immediately following a terrorist attack (event times +1 and +2). This finding suggests that citizens may initially turn toward the military as a source of protection, but this confidence quickly disappears. Consequently, the temporary increase in support for military rule observed earlier does not reflect a structural shift in political preferences but rather a momentary reaction driven by the need for security.

Finally, although Afrobarometer also includes a question asking respondents to evaluate how the government is handling security issues related to extremist groups, this question was only included in one survey round. As a result, the number of available observations is limited, and the estimated coefficients are statistically insignificant, although pointing in the right direction. 

\subsection{Impact on Economic \& Living condition Hardship}

We now turn to the third mechanism, which explores the economic consequences of terrorism. Beyond shaping political attitudes and institutional trust, terrorist activity can disrupt local economies, limit mobility, and undermine livelihoods. To capture these effects, we examine how exposure to terrorist attacks influences economic activity.

We use nighttime light intensity as a proxy for economic activity, drawing on satellite-based nightlight data that reflect the spatial distribution of production and infrastructure use. In addition, we rely on survey-based measures from Afrobarometer to compute the living poverty index and unemployment status. 

\begin{figure}[ht]
    \centering
    \includegraphics[width=1\textwidth]{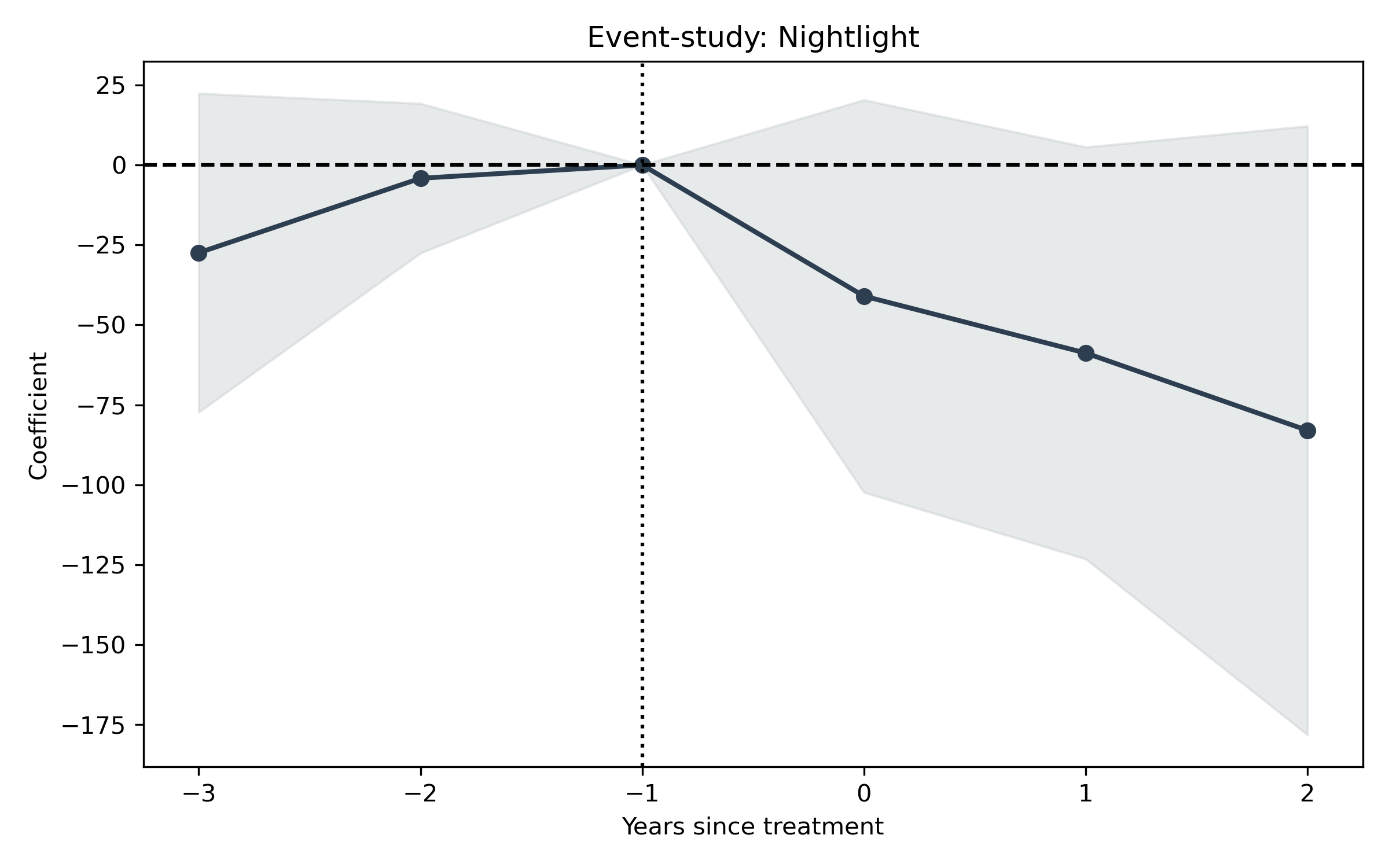}
    \caption{Effect of terrorism on economic activity as measured by nightlight time}
    \label{ntl_sum}
\end{figure}

The results presented in Figure \ref{ntl_sum} and Table \ref{table3d} highlight the negative economic effects of terrorist attacks. The event-study coefficients on nightlight intensity indicate a steady decline in local economic activity following exposure to terrorism. While the coefficients before the attack are close to zero, suggesting no pre-trend, nightlight levels fall in the years after treatment.

Consistent with this evidence, Table \ref{table3d} shows that terrorist violence significantly worsens living conditions. The coefficient on unemployment becomes negative and significant two years after an attack, suggesting a deterioration in labor market opportunities.

Similarly, the Living Poverty Index (LPI) rises by 0.09 points on average after exposure; however, the effect does not appear significant.
\begin{table}[ht!]
\centering
\begin{threeparttable}
\renewcommand{\tablename}{Table}

\caption{Trade-off between Security \& Freedom}
\label{table3d}

\begin{tabular}{lcc}
\hline\hline
            & (1) & (2)  \\
            & Unemployed & LPI \\
\hline
Treated $\times$ Post&        $0.32^{**}$ &        0.09         \\
            &      (0.14)         &      (0.24)         \\
\hline
R-squared   &        0.19         &        0.18         \\
N           &    11,186       &    11,191       \\
\hline
Sub-experiment $\times$ Departments FEs & Yes & Yes \\
Sub-experiment $\times$ Event-time FEs  & Yes & Yes \\
Attacks types FEs                       & Yes & Yes \\
Individual controls                     & Yes & Yes \\
Departments covariates                  & Yes & Yes \\
\hline\hline
\end{tabular}

\begin{tablenotes}
  \footnotesize
  \item Standard errors in parentheses are clustered at the department level. $^{*} p<0.05$, $^{**} p<0.01$, $^{***} p<0.001$.
\end{tablenotes}

\end{threeparttable}
\end{table}

\section{Conclusion}
This paper provides new empirical evidence on how chronic terrorism reshapes political preferences in fragile democracies. Using a stacked Difference-in-Differences framework applied to multiple Afrobarometer rounds (2008–2022) and geocoded terrorism data from ACLED and GTD, we identify the causal effect of successful terrorist attacks on citizens' support for democracy and military rule in Burkina Faso. 

Our results reveal that successful attacks significantly increase approval of military rule while simultaneously decreasing support for democracy. 
These effects appear one year after an attack and are short-lived. The robustness checks confirm that the observed effects are not driven by compositional bias or model specification. Controlling for spillovers across neighboring departments further reinforces the internal validity of the findings.
Exploring potential mechanisms, we show that terrorism reshapes preferences through three key channels. First, attacks increase fear and shift citizens' priorities toward security over freedom, illustrating a security–freedom trade-off. Second, terrorism decreases trust in democratic institutions, especially in the presidency and parliament. Third, terrorist violence deteriorates local economic conditions, raising unemployment and poverty.

Taken together, these findings demonstrate that terrorism can generate a short-term authoritarian drift even within societies that have recently adopted democracy. The short-term decline in democratic support suggests that democratic norms are resilient despite the persistence of violence. Thus, a new challenge for fragile democracies arises: improve security while preserving and developing democratic institutions.

\newpage
\printbibliography[heading = bibintoc, title={Bibliography}]

\newpage
\appendix
\renewcommand{\thetable}{C\arabic{table}}
\renewcommand{\thefigure}{B\arabic{figure}}

\section{ACLED Classification Procedure}
\label{appendix:classification}
This appendix describes the whole procedure used to classify successful terrorism events from the ACLED dataset.
\paragraph{Identifying terrorist attack :} To identify terrorist attacks within the ACLED event dataset, we classified each event as either a terrorist attack or not. The ACLED database reports the names of the two leading actors involved in each event \footnote{Actors can be state forces, rebels, militias, demonstrators, civilians, or external forces}. Based on this information, we first created a variable, \textit{jihadist}, to flag events involving terrorist groups. Since the data do not specify aggressors or targets, we relied on additional indicators.
Events identified by ACLED as targeting civilians and involving jihadist actors were coded as terrorist attacks. For other events, we used the sub-event type (a more granular categorization) and the descriptive notes variable to refine classification.\footnote{Sub-event types include, among others, abduction, armed clash, Looting, suicide bomb, and remote explosive/IED.} Combining these sources allowed us to distinguish terrorist attacks from other forms of violence accurately. After classification, we identify 8,037 events involving a terrorist group and 5,872 terror attacks. 

A key difference between our classification of ACLED data and the GTD database concerns the treatment of \textit{looting/property destruction events}. In Burkina Faso, such incidents are often carried out by armed groups to instill fear, assert territorial control, and destabilize communities. These acts go beyond ordinary criminal behavior and form part of broader insurgent strategies aimed at undermining state authority and public security. Consequently, we classify looting/property destruction as acts of terrorism. We replicate the analysis, not using looting/property destruction as attacks, to validate the robustness of our results, with corresponding tables reported in the Appendix.

\paragraph{Identifying successful attack:} 
We construct a variable, \textit{successful}, indicating whether a terrorist attack was successful (1) or failed (0). Our identification strategy relies on this variable. Success criteria vary by attack type.\footnote{An assassination is successful if the target is killed; an explosion if the device detonates; a hijacking or kidnapping if the perpetrators gain control of the vehicle or victim; an armed assault if the target (person or property) is struck; and an infrastructure attack if the facility is damaged. An unarmed assault is deemed successful if it causes injury.} For looting or property destruction, only events involving actual building destruction are classified as successful; mere looting is not.\footnote{The following interactive dashboard provides a visual representation of key insights related to successful terrorism incidents and agricultural distribution in Burkina Faso. Users can explore the data dynamically by filtering different years and provinces. Access the dashboard here: \href{https://public.tableau.com/app/profile/p.carmel.marie.frederique.zagre/viz/BF_17424541824390/Dashboard1?publish=yes}{Tableau Dashboard}.}

There are 69 terrorist attacks for which the success status could not be determined, as the event descriptions only specify an unknown outcome. These events were dropped from the analysis. All terrorist attacks are aggregated at the department-year level. If at least one terrorist attack in a given department-year cell is classified as successful, the variable \textit{Successful} takes the value of one.

Between 2015 and June 23, 2021, the GTD reports 444 terrorist attacks, of which 411 were successful, corresponding to a success rate of 92.6\%.

Using data from the ACLED and excluding looting incidents from the definition of terrorist attacks, we identify 1,757 attacks between 2015 and the end of 2021, 958 of which were successful (54.5\% success rate). When restricting the ACLED sample to the same time window as the GTD (2015–June 23, 2021), we record 1,283 terrorist attacks, among which 756 were successful, yielding a success rate of 58.9\%.

\medskip
\noindent\textit{Note:} The difference in the total number of attacks and in success rates between ACLED and GTD (58.9\% vs.\ 92.6\%) reflects differences in event classification, but also time coverage. The GTD records only incidents meeting strict terrorism criteria and is available up to mid-2021, whereas our classification includes a broader range of political violence events and capture event up until 2022. Given our identification strategy relies on within-department variation in attack success rather than on absolute counts, and because ACLED provides more  recent data for Burkina Faso, we use ACLED as the primary source. Results based on the GTD are presented in Appendix \ref{appendix:tables} in Table \ref{table2c} and show qualitatively similar patterns.

\newpage
\section{Additional Figures}
\label{appendix:fig}

\begin{figure}[ht]
    \centering
    \begin{subfigure}[t]{0.48\textwidth}
        \centering
        \includegraphics[width=\textwidth]{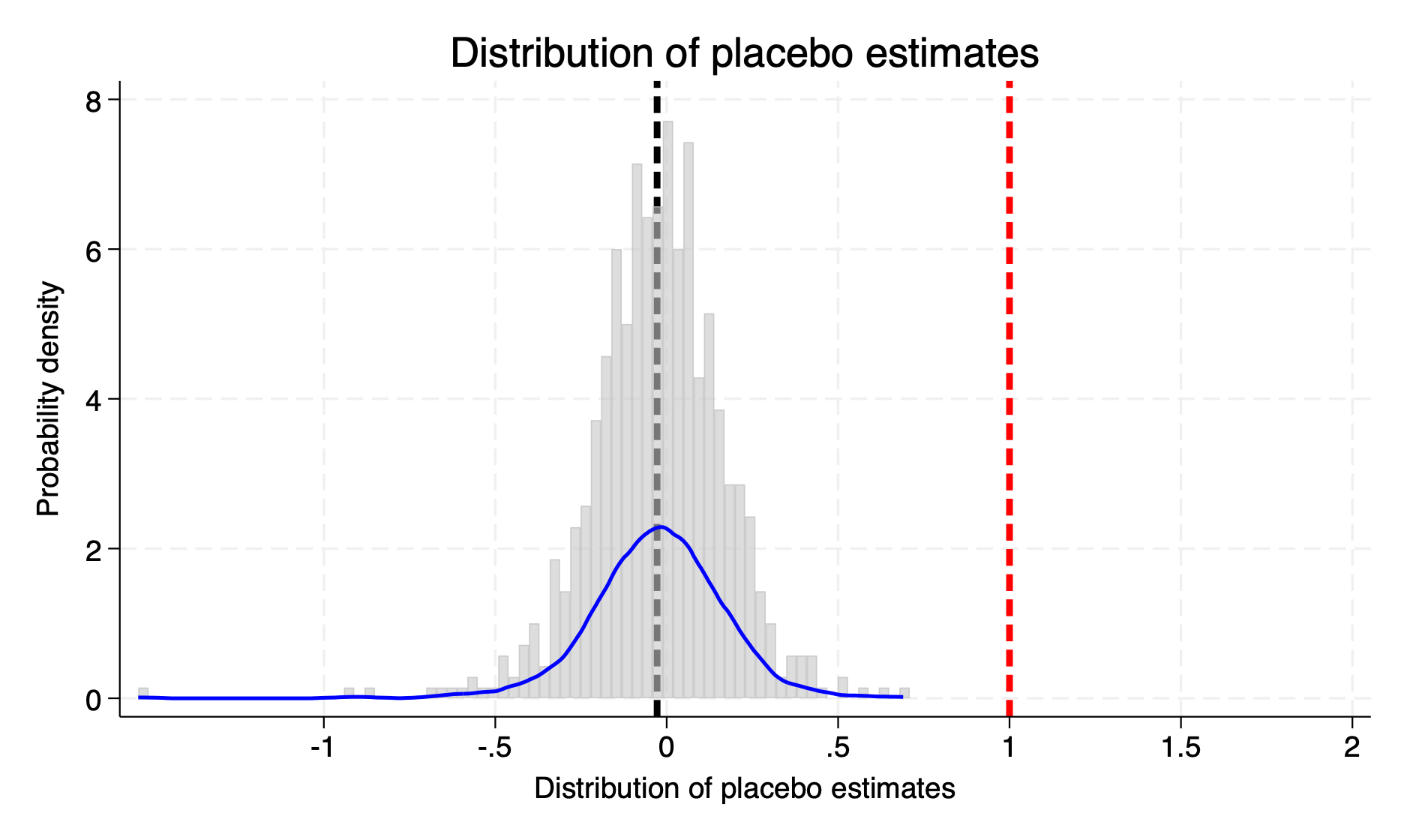}
        \caption{Placebo test for support for military rule}
        \label{fig:placebo_military}
    \end{subfigure}
    \hfill
    \begin{subfigure}[t]{0.48\textwidth}
        \centering
        \includegraphics[width=\textwidth]{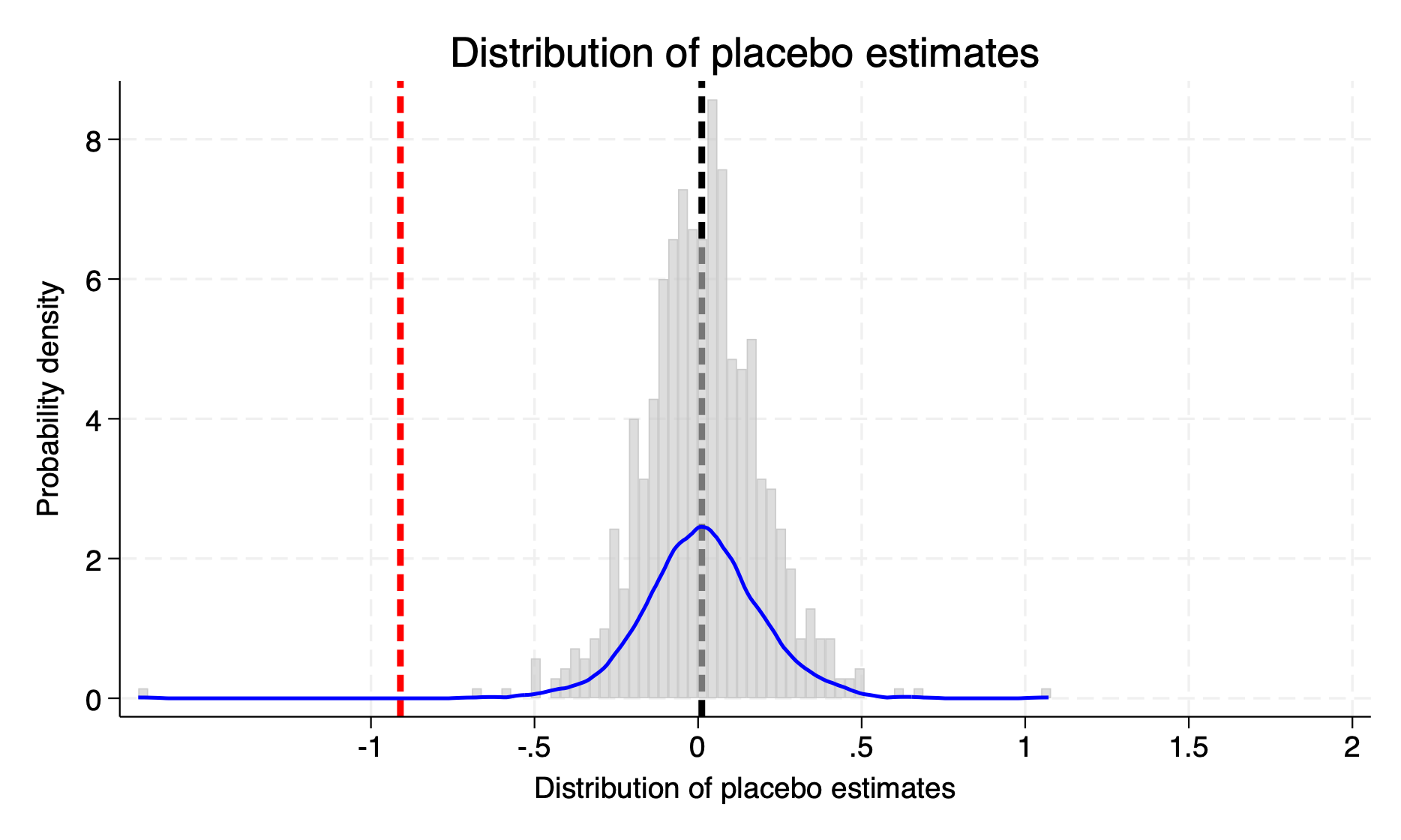}
        \caption{Placebo test for support for democracy}
        \label{fig:placebo_democracy}
    \end{subfigure}
    \caption{Permutation (placebo) tests showing the distribution of estimated ATT under random treatment timing for both outcomes.}
    \label{fig:placebo_combined}
\end{figure}

\begin{figure}[ht]
    \centering
    \begin{subfigure}[t]{0.32\textwidth}
        \centering
        \includegraphics[width=\textwidth]{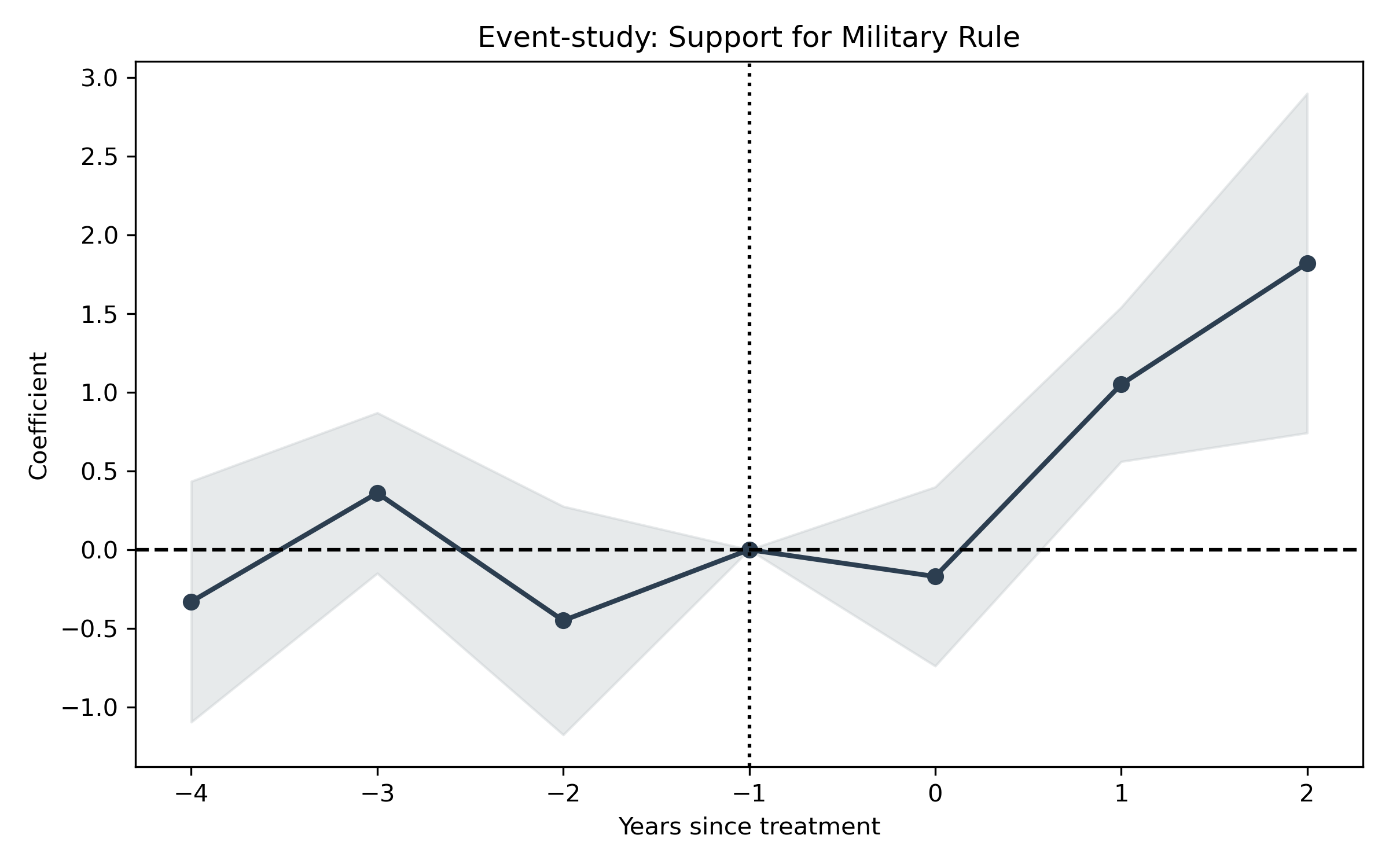}
        \caption{Window [4,2]}
    \end{subfigure}
    \hfill
    \begin{subfigure}[t]{0.32\textwidth}
        \centering
        \includegraphics[width=\textwidth]{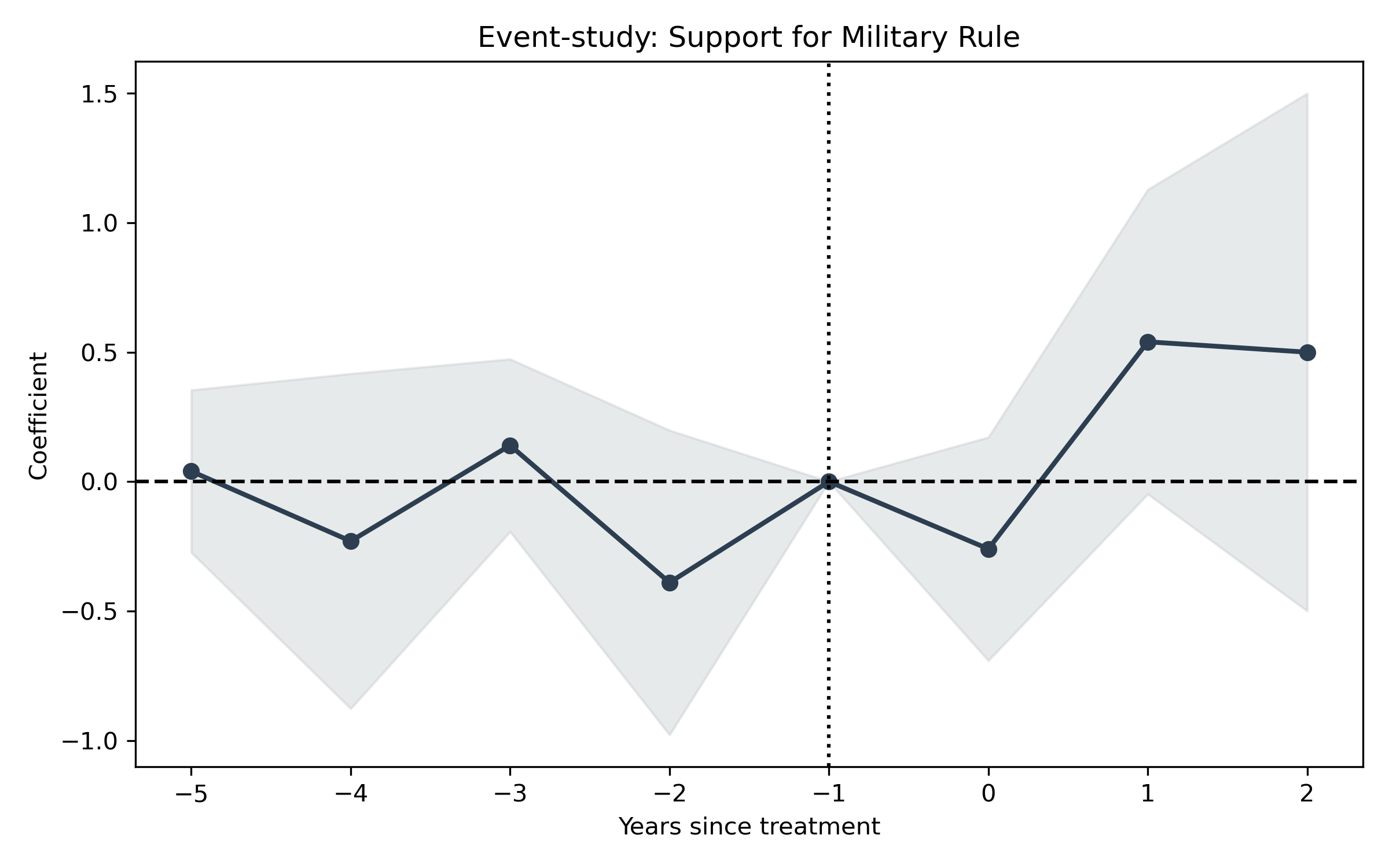}
        
        \caption{Window [5,2]}
    \end{subfigure}
    \hfill
    \begin{subfigure}[t]{0.32\textwidth}
        \centering
        \includegraphics[width=\textwidth]{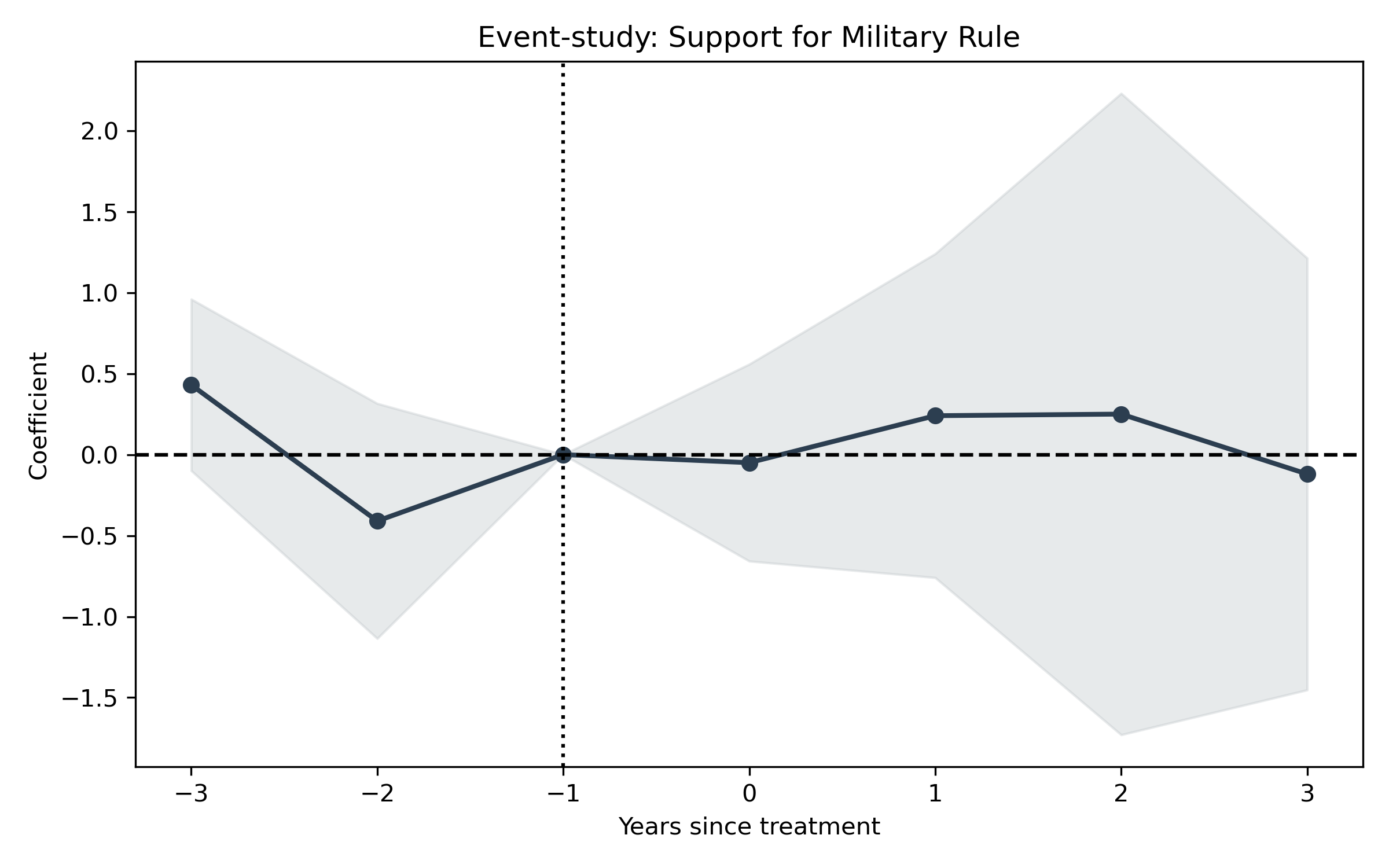}

        \caption{Window [3,3]}
    \end{subfigure}
    
    \vspace{0.5em}
    \begin{subfigure}[t]{0.32\textwidth}
        \centering
        \includegraphics[width=\textwidth]{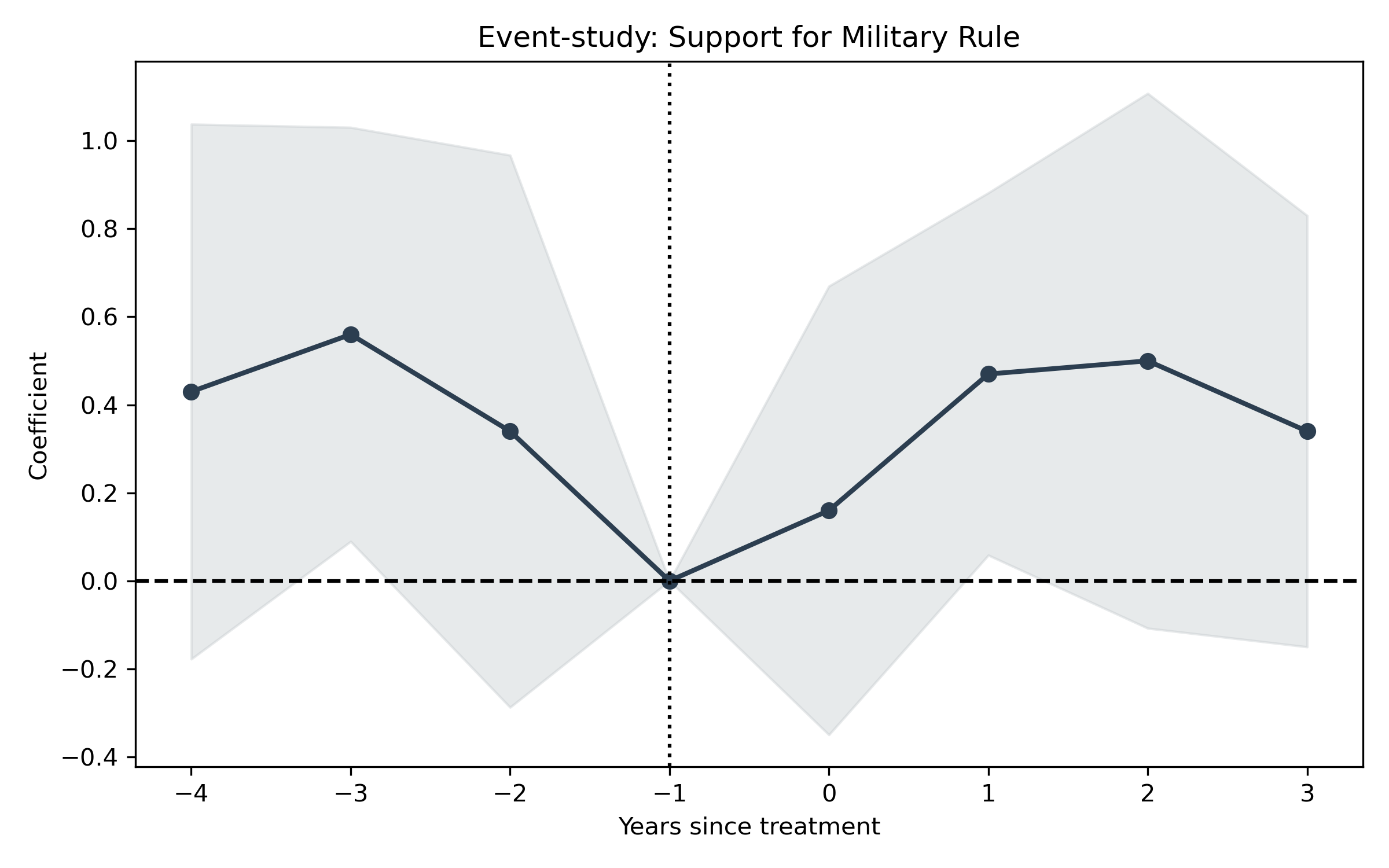}

        \caption{Window [4,3]}
    \end{subfigure}

    \caption{Event-study estimates for support for military rule under alternative time windows. Each panel reports coefficients from stacked DiD models using different pre- and post-treatment horizons.}
    \label{fig:windows_combined_military}
\end{figure}

\begin{figure}[ht]
    \centering
    \begin{subfigure}[t]{0.32\textwidth}
        \centering
        \includegraphics[width=\textwidth]{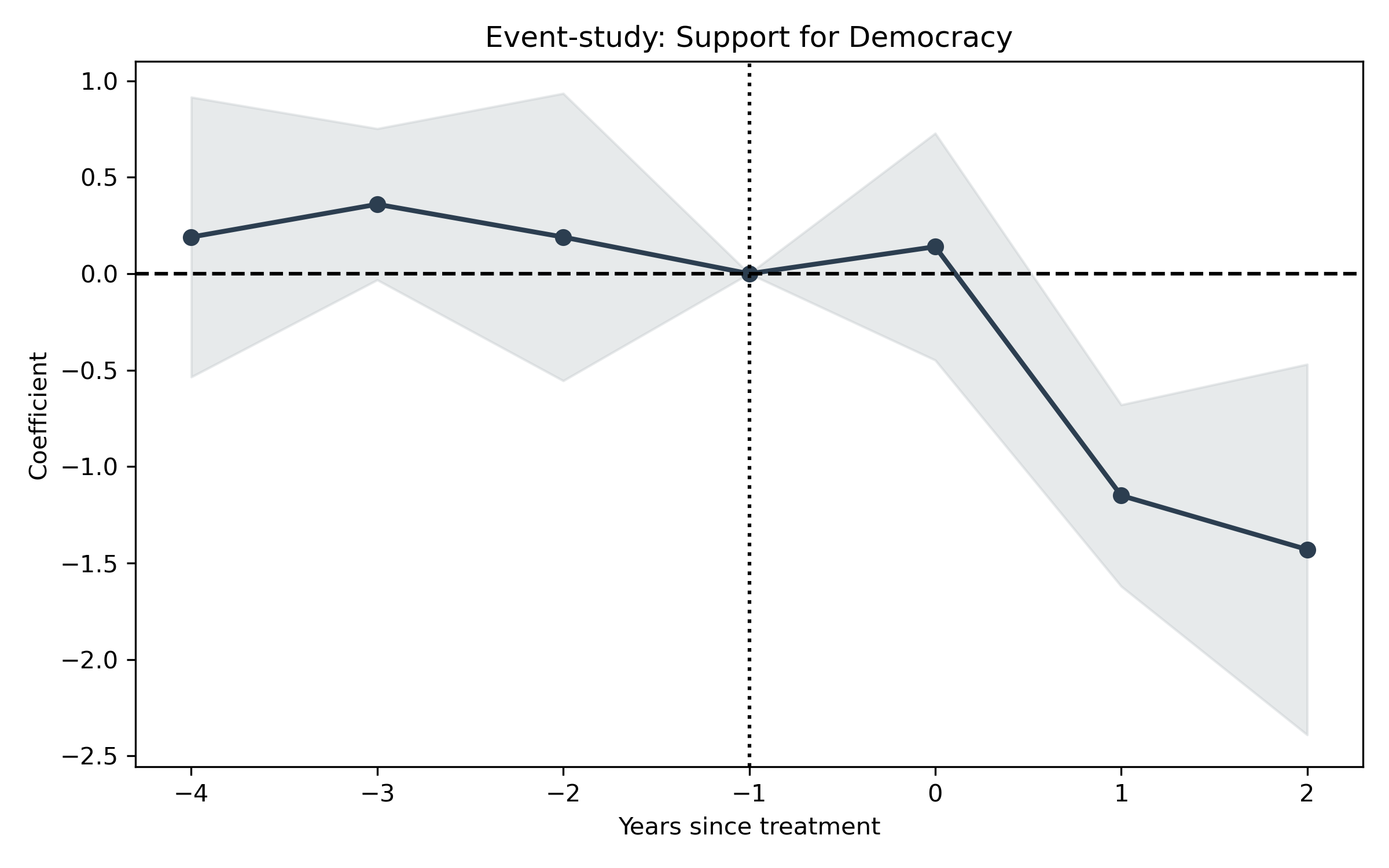}

        \caption{Window [4,2]}
    \end{subfigure}
    \hfill
    \begin{subfigure}[t]{0.32\textwidth}
        \centering
        \includegraphics[width=\textwidth]{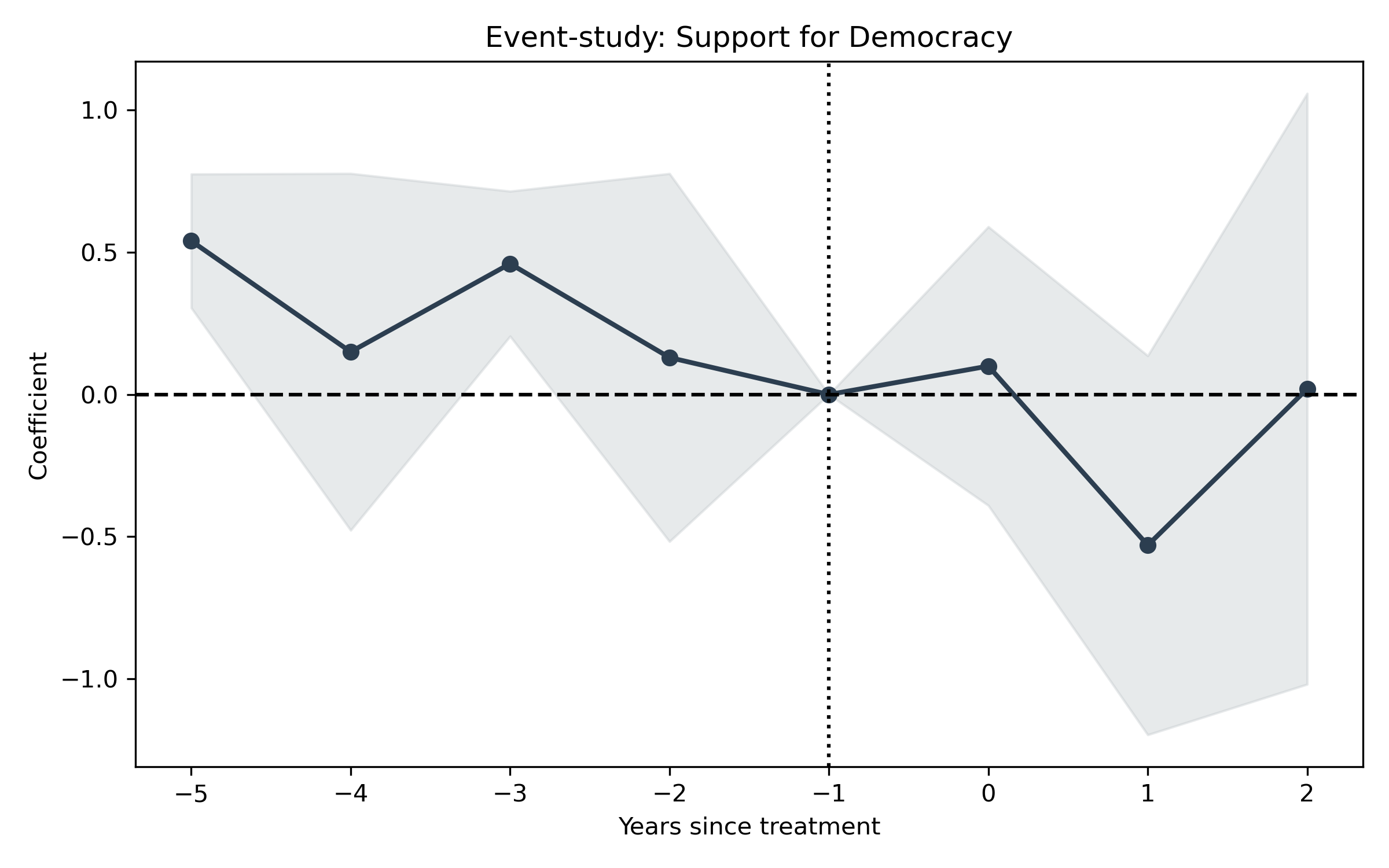}
        \caption{Window [5,2]}
    \end{subfigure}
    \hfill
    \begin{subfigure}[t]{0.32\textwidth}
        \centering
        \includegraphics[width=\textwidth]{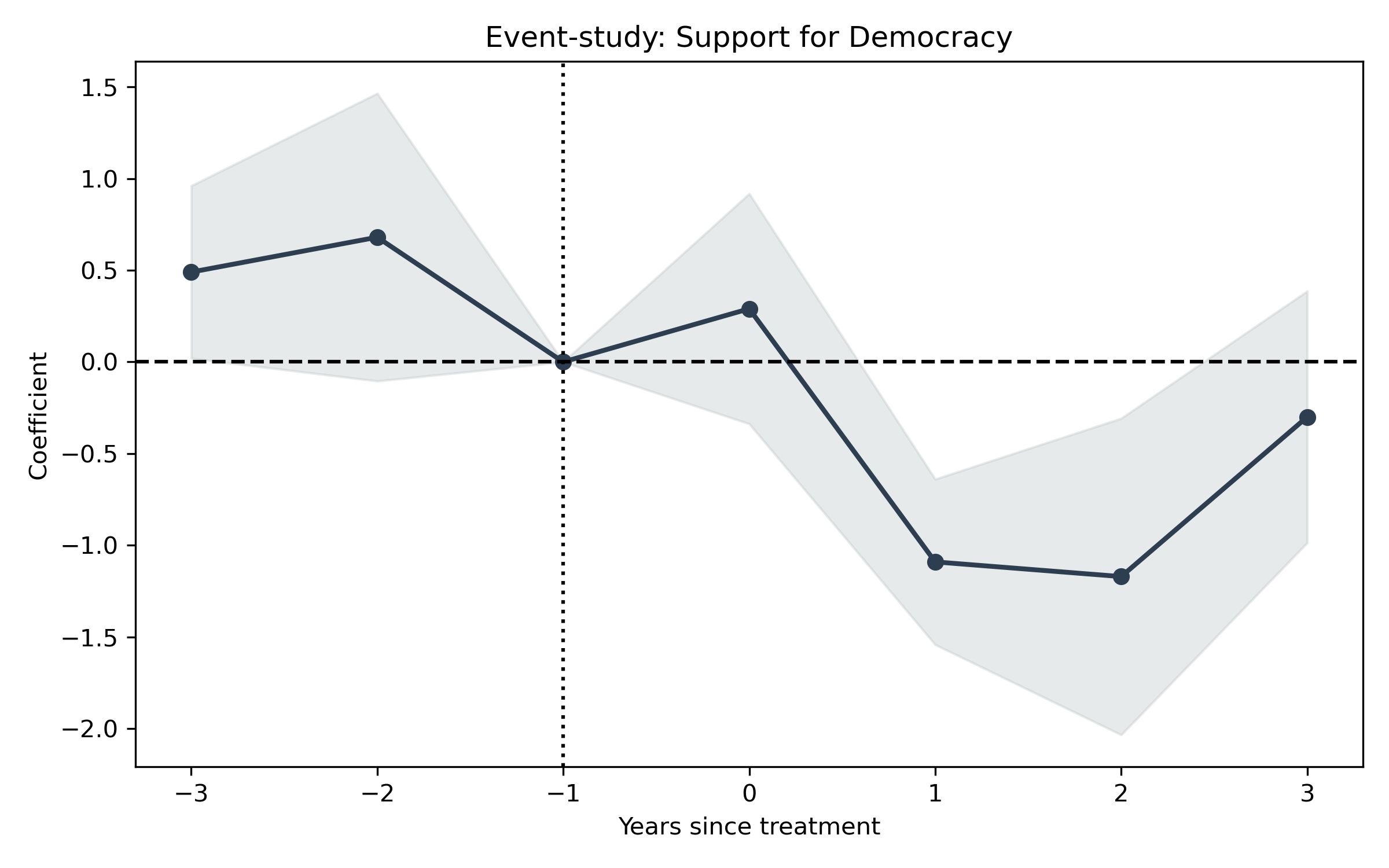}
        \caption{Window [3,3]}
    \end{subfigure}
    
    \vspace{0.5em}
    \begin{subfigure}[t]{0.32\textwidth}
        \centering
        \includegraphics[width=\textwidth]{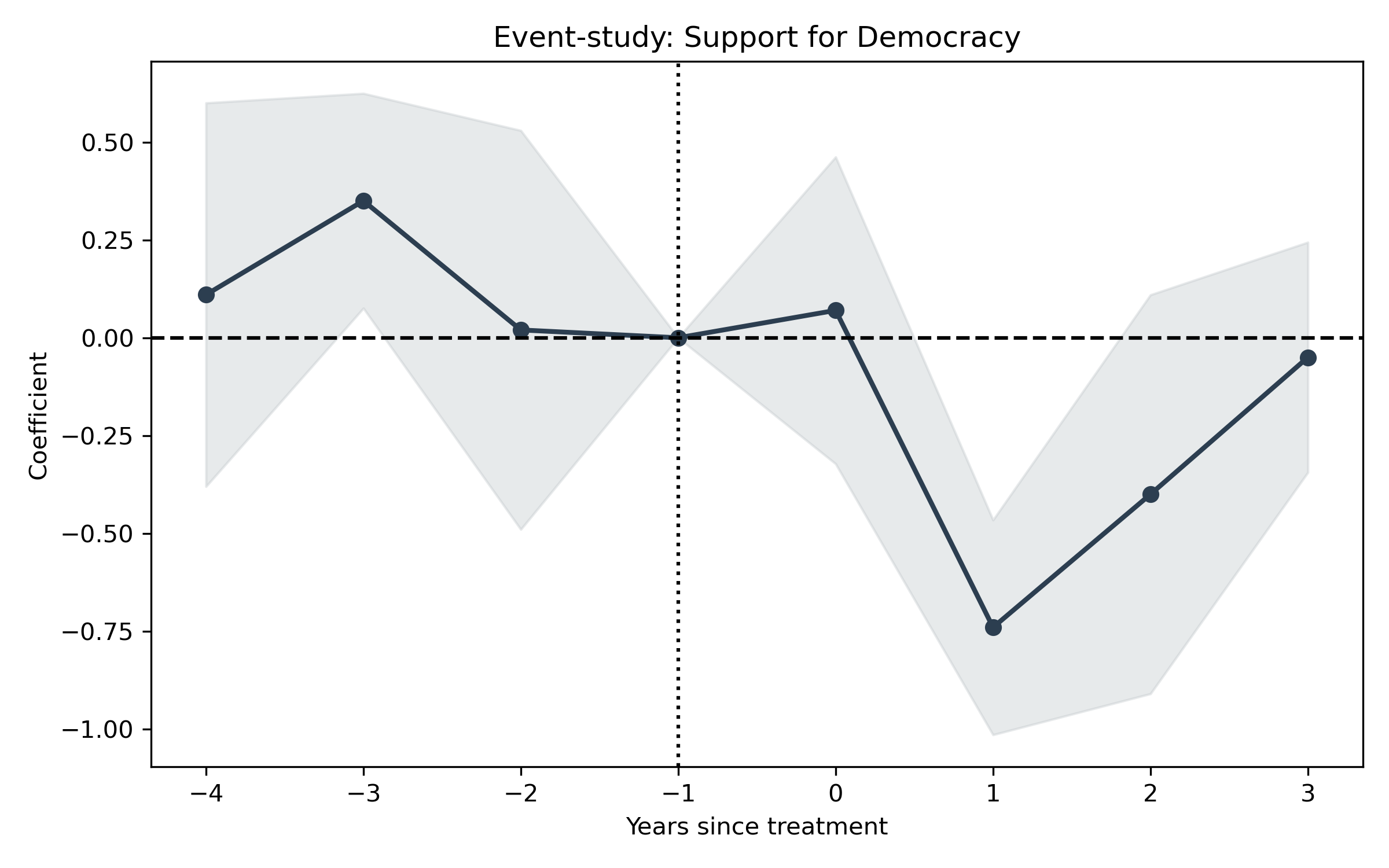}
        \caption{Window [4,3]}
    \end{subfigure}

    \caption{Event-study estimates for support for democratic rule under alternative time windows. Each panel reports coefficients from stacked DiD models using different pre- and post-treatment horizons.}
    \label{fig:windows_combined_dem}
\end{figure}

\newpage
\section{Additional Tables}
\label{appendix:tables}
\begin{table}[ht]
\centering
\caption{Data Sources and Variables Used}
\label{tab:data_sources}
\begin{tabular}{p{6cm} p{6cm}}
\toprule
\textbf{Source} & \textbf{Description}  \\
\midrule

\textbf{Institut National de la Statistique et de la Démographie}
& Statistical agency of Burkina Faso \\\
& \\
\textbf{Supernum Dataset}
&SUPERMUN is an annual department-level dataset that reports standardized indicators of institutional capacity and public service delivery for all communes in Burkina Faso.\\\
& \\

\textbf{AidData}
& Geospatial and environmental indicators derived from satellite data. \\\
& \\

\textbf{Humanitarian Data Exchange (HDX)}
& Open humanitarian and geographic datasets. \\

\bottomrule
\end{tabular}
\end{table}

\begin{table}[ht!]
\centering
\begin{threeparttable}
\renewcommand{\tablename}{Table}

\caption{Successful Terror Attacks and Political Attitudes – Never-Treated Control Group}
\label{table2j}

\begin{tabular}{lcc}
\hline\hline
            & (1) & (2) \\
            & Military rule & Support democracy \\
\hline            
Treated $\times$ Event-time, -3&       $ 0.85^{**}$ &        0.52         \\
            &      (0.29)         &      (0.36)         \\
[1em]
Treated $\times$ Event-time, -2&       -0.01         &        0.91         \\
            &      (0.41)         &      (0.46)         \\
[1em]
Treated $\times$  Event-time, 0&        0.40         &        0.45         \\
            &      (0.33)         &      (0.39)         \\
[1em]
Treated $\times$  Event-time, +1&        $0.97^{***}$&       $-1.34^{***}$\\
            &      (0.27)         &      (0.24)         \\
[1em]
Treated $\times$  Event-time, +2&        $2.17^{***}$&       $-1.47^{*} $ \\
            &      (0.63)         &      (0.63)         \\
\hline
Treated $\times$ Post&$1.18^{***}  $       &$-0.78^{*}$         \\
            &        (0.34)         &        (0.36)         \\
R-Squared   &        0.15         &        0.14         \\
N           &     8,064         &     7,981         \\

\hline
Sub-experiment $\times$ Departments FEs & Yes & Yes \\
Sub-experiment $\times$ Event-time FEs  & Yes & Yes \\
Attacks types FEs                       & Yes & Yes \\
Individual controls                     & Yes & Yes \\
Departments covariates                  & Yes & Yes \\
\hline\hline
\end{tabular}

\begin{tablenotes}
  \footnotesize
  \item Standard errors in parentheses are clustered at the department level. $^{*} p<0.05$, $^{**} p<0.01$, $^{***} p<0.001$. 
\end{tablenotes}

\end{threeparttable}
\end{table}

\begin{table}[ht!]
\centering
\begin{threeparttable}
\renewcommand{\tablename}{Table}

\caption{Successful Terror Attacks and Political Attitudes – Entropy Balance Matching}
\label{table4a}

\begin{tabular}{lcc}
\hline\hline
            & (1) & (2) \\
            & Military rule & Support democracy \\
\hline            
Treated $\times$ Event-time, -3&        0.24         &        0.41         \\
            &      (0.22)         &      (0.24)         \\
[1em]
Treated $\times$ Event-time, -2&       $-1.02^{***}$&       $ 1.00^{*} $ \\
            &      (0.29)         &      (0.46)         \\
[1em]
Treated $\times$  Event-time, 0&       -0.44         &        0.45         \\
            &      (0.24)         &      (0.31)         \\
[1em]
Treated $\times$  Event-time, +1&        $1.25^{***}$&       $-1.48^{***}$\\
            &      (0.26)         &      (0.29)         \\
[1em]
Treated $\times$  Event-time, +2&       $ 1.96^{***}$&       $-1.84^{***}$\\
            &      (0.54)         &      (0.49)         \\
\hline
Treated $\times$ Post&$0.92^{**}$         &$-0.96^{***}$         \\
            &        (0.29)         &        (0.27)         \\
R-Squared   &        0.16         &        0.15         \\
N           &    11,098         &    10,979       \\

\hline
Sub-experiment $\times$ Departments FEs & Yes & Yes \\
Sub-experiment $\times$ Event-time FEs  & Yes & Yes \\
Attacks types FEs                       & Yes & Yes \\
Individual controls                     & Yes & Yes \\
Departments covariates                  & Yes & Yes \\
\hline\hline
\end{tabular}

\begin{tablenotes}
  \footnotesize
  \item Standard errors in parentheses are clustered at the department level. $^{*} p<0.05$, $^{**} p<0.01$, $^{***} p<0.001$. 
\end{tablenotes}

\end{threeparttable}
\end{table}

\begin{table}[ht!]
\centering
\begin{threeparttable}
\renewcommand{\tablename}{Table}

\caption{Successful Terror Attacks and Political Attitudes – \textcite{gtd2021}}
\label{table2c}

\begin{tabular}{lcc}
\hline\hline
            & (1) & (2) \\
            & Military rule & Support democracy \\
\hline
Treated $\times$ Event-time, -3&       -0.14         &        0.07         \\
           &      (0.08)         &      (0.15)         \\
[1em]
Treated $\times$ Event-time, -2&       $ 0.69^{**}$ &       -0.23         \\
           &      (0.23)         &      (0.13)         \\
[1em]
Treated $\times$  Event-time, 0&        $0.91^{***}$&       $-0.46^{***}$\\
           &      (0.18)         &      (0.12)         \\
[1em]
Treated $\times$  Event-time, +1&       $-0.50^{***}$&        0.21         \\
           &      (0.09)         &      (0.19)         \\
[1em]
Treated $\times$  Event-time, +2&        $0.45^{***}$&       $-0.28^{**} $\\
           &      (0.08)         &      (0.09)         \\
\hline
Treated $\times$ Post&$0.29^{***} $        &$-0.18^{*} $        \\
           &        (0.08)         &        (0.09)         \\
R-Squared   &        0.15         &        0.19         \\
N           &    14, 849    &    14, 692       \\

\hline
Sub-experiment $\times$ Departments FEs & Yes & Yes \\
Sub-experiment $\times$ Event-time FEs  & Yes & Yes \\
Attacks types FEs                       & Yes & Yes \\
Individual controls                     & Yes & Yes \\
Departments covariates                  & Yes & Yes \\
\hline\hline
\end{tabular}

\begin{tablenotes}
  \footnotesize
  \item Standard errors in parentheses are clustered at the department level. $^{*} p<0.05$, $^{**} p<0.01$, $^{***} p<0.001$. 
\end{tablenotes}

\end{threeparttable}
\end{table}

\begin{table}[ht!]
\centering
\begin{threeparttable}
\renewcommand{\tablename}{Table}

\caption{Successful Terror Attacks and Political Attitudes – Without Looting}
\label{table2b}

\begin{tabular}{lcc}
\hline\hline
            & (1) & (2) \\
            & Military rule & Support democracy \\
\hline

Treated $\times$ Event-time, -3&        0.03         &        0.23         \\
            &      (0.24)         &      (0.19)         \\
[1em]
Treated $\times$ Event-time, -2&       -0.21         &       -0.11         \\
            &      (0.27)         &      (0.23)         \\
[1em]
Treated $\times$  Event-time, 0&        0.12         &       -0.47         \\
            &      (0.31)         &      (0.27)         \\
[1em]
Treated $\times$  Event-time, +1&        $1.14^{***}$&       $-1.18^{***}$\\
            &      (0.26)         &      (0.12)         \\
[1em]
Treated $\times$  Event-time, +2&        0.16         &       -0.12         \\
            &      (0.37)         &      (0.28)         \\
\hline
Treated $\times$ Post& $ 0.48^{}  $       &$-0.59^{**} $        \\
            &       ( 0.24)         &        (0.20 )        \\
R-Squared   &        0.14         &        0.14         \\
N           &    11,906     &    11,791    \\

\hline
Sub-experiment $\times$ Departments FEs & Yes & Yes \\
Sub-experiment $\times$ Event-time FEs  & Yes & Yes \\
Attacks types FEs                       & Yes & Yes \\
Individual controls                     & Yes & Yes \\
Departments covariates                  & Yes & Yes \\
\hline\hline
\end{tabular}

\begin{tablenotes}
  \footnotesize
  \item Standard errors in parentheses are clustered at the department level. $^{*} p<0.05$, $^{**} p<0.01$, $^{***} p<0.001$. 
\end{tablenotes}

\end{threeparttable}
\end{table}

\begin{table}[htbp]
\centering
\caption{Treated Departments by Sub-experiment (ACLED vs GTD)}
\resizebox{\textwidth}{!}{%
\begin{tabular}{|c|p{9cm}|p{9cm}|}
\hline
\textbf{Sub-experiment} & \textbf{ACLED (Treated)} & \textbf{GTD (Treated)} \\ \hline
2015 & Tongomayel & Bagassi, Ouagadougou, Samôgôgouan, Pissila \\ \hline
2016 & Ouagadougou & Djibo, Dori \\ \hline
2017 & Barani, Arbinda, Baraboulé, Diguel, Djibo & Djibasso, Arbinda, Baraboulé \\ \hline
2018 & Bourzanga, Fada-Ngourma, Matiakoali, Satiri, Pama, Soudougui, Titao, Gorom-Gorom, Kampti, Pissila, Dori, Partiaga, Mansila, Solhan &
Bourzanga, Bogandé, Mani, Fada-Ngourma, Matiakoali, Pama, Barani, Soudougui, Titao, Gorom-Gorom, Lanfièra, Partiaga, Sebba, Solhan \\ \hline
2019 & Rollo, Bitou, Niangoloko, Sidéradougou, Diapangou, Yamba, Ouindigui, Tougouri, Nako, Gorgadji, Sebba, Koumbri, Senguènèga &
Sanaba, Bitou, Niangoloko, Ouindigui, Tougouri, Bani, Gorgadji, Mansila, Koumbri, Ouahigouya \\ \hline
2020 & Sanaba, Mangodara, Bilanga, Mani, Tougan, Bani, Falagountou, Botou, Titabè, Ouahigouya, Goursi &
Bilanga, Boussouma, Tougan \\ \hline
2021 & Nouna, Dédougou, Pouni & Banfora \\ \hline
2022 & Pâ, Solenzo, Zabré, Bogandé, Tibga, Dandé, Lalgaye, Kourouma, Bondokui, Yé, Dassa, Ténado, Mané, Founzan, Houndé, Lèba, Gogo & NA \\ \hline
\end{tabular}%
}
\label{tab:treated_only}
\end{table}

\end{document}